\documentclass[aps,onecolumn,superscriptaddress,amsmath,amssymb,pre,floatfix,10pt]{revtex4-1}

\usepackage{hyphenat}
\usepackage{hhline}
\usepackage{upgreek}
\usepackage{graphicx}
\usepackage[table]{xcolor}
\newcommand{\ATPfrac}{\ensuremath{\alpha_{\mathrm{ATP}}}}

\newcommand{\rmlabels}[3]{\ensuremath{#1^{\mathrm{#2}}_{\mathrm{#3}}}}
\newcommand{\brmlabels}[3]{\ensuremath{{\bf #1}^{\mathrm{#2}}_{\mathrm{#3}}}}

\def\longrightharpoonup{\relbar\joinrel\rightharpoonup}
\def\longleftharpoondown{\leftharpoondown\joinrel\relbar}

\def\longrightleftharpoons{
  \mathop{
    \vcenter{
      \hbox{
      \ooalign{
        \raise1pt\hbox{$\longrightharpoonup\joinrel$}\crcr
	  \lower1pt\hbox{$\longleftharpoondown\joinrel$}
	  }
      }
    }
  }
}

\newcommand{\rates}[2]{\displaystyle
  \mathrel{\longrightleftharpoons^{#1\mathstrut}_{#2}}}

\newcommand{\tref}[1]{Table~\ref{table:#1}}
\newcommand{\trefs}[1]{Tables~\ref{table:#1}}
\newcommand{\tlabel}[1]{\label{table:#1}}

\newcommand{\fref}[1]{Fig.~\ref{fig:#1}}

\newcommand{\frefs}[1]{Figs.~\ref{fig:#1}}
\newcommand{\flabel}[1]{\label{fig:#1}}

\newcommand{\eref}[1]{Eq.~\ref{eqn:#1}}

\newcommand{\erefstwo}[2]{Eqs.~\ref{eqn:#1}~and~\ref{eqn:#2}}

\newcommand{\elabel}[1]{\label{eqn:#1}}

\begin{document}

\title{A Thermodynamically consistent model of the post-translational Kai circadian clock}

\author{Joris \surname{Paijmans}}
\affiliation{FOM Institute AMOLF, Science Park 104, 1098 XG Amsterdam, The Netherlands}
\author{David K. \surname{Lubensky}}
\affiliation{Department of Physics, University of Michigan, Ann Arbor, MI 48109-1040, USA}
\author{Pieter Rein \surname{ten Wolde}}
\affiliation{FOM Institute AMOLF, Science Park 104, 1098 XG Amsterdam, The Netherlands}

\begin{abstract}
 The principal pacemaker of the circadian clock of the cyanobacterium
  {\it S. elongatus} is a protein phosphorylation cycle consisting of
  three proteins, KaiA, KaiB and KaiC. KaiC forms a homohexamer, with
  each monomer consisting of two domains, CI and CII. Both domains can
  bind and hydrolyze ATP, but only the CII domain can be
  phosphorylated, at two residues, in a well-defined sequence.  While
  this system has been studied extensively, how the clock is driven
  thermodynamically has remained elusive.  Inspired by recent
  experimental observations and building on ideas from previous mathematical
  models, we present a new, thermodynamically consistent, 
  statistical-mechanical model of the clock. 
  At its heart are two main ideas: $i$) ATP hydrolysis in
  the CI domain provides the thermodynamic driving force for the
  clock, switching KaiC between an active conformational state in
  which its phosphorylation level tends to rise and an inactive one
  in which it tends to fall; $ii$) phosphorylation of the CII domain
  provides the timer for the hydrolysis in the CI domain. The model
  also naturally explains how KaiA, by acting as a nucleotide exchange
  factor, can stimulate phosphorylation of KaiC, and how the
  differential affinity of KaiA for the different KaiC phosphoforms
  generates the characteristic temporal order of KaiC
  phosphorylation. As the phosphorylation level in the CII domain
  rises, the release of ADP from CI slows down, making the inactive
  conformational state of KaiC more stable. In the inactive state,
  KaiC binds KaiB, which not only stabilizes this state further, but
  also leads to the sequestration of KaiA, and hence to KaiC
  dephosphorylation. Using a dedicated kinetic Monte Carlo algorithm,
  which makes it possible to efficiently simulate this system
  consisting of more than a billion reactions, we show that the model
  can describe a wealth of experimental data.  
\end{abstract}

\maketitle

% Please keep the Author Summary between 150 and 200 words
% Use first person. PLOS ONE authors please skip this step. 
% Author Summary not valid for PLOS ONE submissions. 
\section*{Author Summary}
Circadian clocks are biological timekeeping devices with a rhythm of 24 hours in living cells 
pertaining to all kingdoms of life. 
They help organisms to coordinate their behavior with the day-night cycle. 
The circadian clock of the cyanobacterium {\it Synechococcus elongatus} 
is one of the simplest and best characterized clocks in biology. 
The central clock component is the protein KaiC, 
which is phosphorylated and dephosphorylated in a cyclical manner with a 24 hr period. 
While we know from elementary thermodynamics that oscillations require a net turnover 
of fuel molecules, in this case ATP, 
how ATP hydrolysis drives the clock has remained elusive. 
Based on recent experimental observations and building on ideas from existing models, 
we construct the most detailed mathematical model of this system to date. 
KaiC consists of two domains, CI and CII, which each can bind ATP, 
yet only CII can be phosphorylated. 
Moreover, KaiC can exist in two conformational states, 
an active one in which the phosphorylation level tends to rise, 
and an inactive one in which it tends to fall. 
Our model predicts that ATP hydrolysis in the CI domain 
is the principal energetic driver of the clock, 
driving the switching between the two conformational states, 
while phosphorylation in the CII domain provides the timer for the conformational switch. 
The coupling between ATP hydrolysis in the CI domain and phosphorylation in the CII domain 
leads to novel testable predictions.

\section*{Introduction}
Circadian clocks, which allow organisms to anticipate changes in day and night, are a fascinating example of biological rhythms. One of
the most studied and best characterized models of circadian oscillations is the
cyanobacterium {\it Synechococcus elongatus}. 
It is now known that it combines a protein synthesis cycle
\cite{Xu2000,Nakahira2004,Nishiwaki2004} with a protein
phosphorylation cycle \cite{Tomita2005}, and in 2005 the latter was
reconstituted in the test tube \cite{Nakajima2005}. 
This stimulated a detailed characterization of its design principles, 
in a fruitful collaboration between experiments and modeling
\cite{Emberly2006,Mehra2006,Rust2007,VanZon2007,Clodong2007,Mori2007,Miyoshi2007,Yoda2007,Eguchi2008,Markson2009,Zwicker2010,Paijmans2016}. 
Yet, it has remained an open question how the clock is driven thermodynamically.

The central components of the protein phosphorylation cycle are three
proteins, KaiA, KaiB, and KaiC. KaiC forms a hexamer with two
phosphorylation sites per monomer, serine 431 and threonine 432
\cite{Nishiwaki2004}, which are phosphorylated and dephosphorylated in
a well-defined temporal order \cite{Nishiwaki2007,Rust2007}.  KaiA
stimulates phosphorylation \cite{Williams2002,Iwasaki2002,Xu2003},
while KaiB negates the effect of KaiA
\cite{Iwasaki2002,Xu2003,Kitayama2003,Pattanayek2008}.  Modeling in
combination with experiments indicate that the oscillations of the
individual hexamers are synchronized via the mechanism of differential
affinity: while KaiA stimulates phosphorylation, the limited supply of
KaiA binds preferentially to those KaiC hexamers that are falling
behind in the cycle, forcing the front runners to slow down and
allowing the laggards to catch up \cite{VanZon2007, Rust2007}.  The
mechanism of differential affinity appears to be active not only
during the phosphorylation phase of the cycle
\cite{VanZon2007,Lin2014}, but also during the dephoshorylation phase,
when KaiC sequesters KaiA via the binding of KaiB \cite{VanZon2007,Rust2007}. 
Monomer exchange between hexamers, observed in experiments \cite{Emberly2006}, 
is an alternative synchronization mechanism.
However, theoretical studies by us and others show that monomer exchange 
is not critical for stable oscillations \cite{Rust2007,VanZon2007,Clodong2007}.

While it is clear that the clock is driven by the turnover of ATP \cite{Terauchi2007, Murakami2008}, 
how fuel turnover drives the phosphorylation oscillations is still unclear. 
In previous models \cite{VanZon2007,Phong2012,Lin2014},
phosphorylation is driven by ATP hydrolysis, 
while dephosphoryation proceeds via the spontaneous release of the phosphate groups from the threonine and serine residues.
Intriguingly, however, recent experiments have revealed that during the dephosphorylation phase of the clock, 
ATP is regenerated \cite{Nishiwaki2012, Egli2012}: the phosphate groups on the serine
and threonine residues are transferred back to ADP. If phosphorylation
and dephosphorylation do not cause any net turnover of ATP, what
then drives the clock? Clocks are necessarily dissipative,
entailing a net turnover of fuel molecules per cycle.

KaiC consists of two highly homologous domains, 
called the CI and the CII domain \cite{Mori2002}. 
Both domains can bind and hydrolyze ATP \cite{Hayashi2003, Hayashi2004}, 
but only the CII domain can be phosphorylated \cite{Nishiwaki2000}. 
The ATP regeneration experiments indicate that the
phosphorylation and dephosphorylation of CII proceeds via the transfer
of phosphate groups between the threonine/serine residues and the nucleotide bound to CII \cite{Nishiwaki2012, Egli2012}, 
leaving open the possibility that there is no net turnover of ATP on the CII domain.  

Here, we argue that the hydrolysis of ATP in the CI domain is the
principal energetic driver of the clock.  We present a new
mathematical model of the post-translational Kai circadian clock in {\it S. elongatus},
which is based on the idea that ATP hydrolysis in CI drives a
conformational switch of KaiC. Previously, it has been predicted
that ATP hydrolysis plays an important role in driving conformational
transitions \cite{Phong2012,Chang2012} and that these transitions are vital to
generating the oscillations \cite{VanZon2007,Lin2014}, 
predictions that have found experimental support
\cite{Kim2008,Chang2011,Chang2012,Phong2012,Egli2013,Snijder2014}. 
Our model, however, goes farther, predicting that ATP hydrolysis in the CI domain is the place where
detailed balance must be broken in order to generate sustained oscillations.
Our model is inspired by that of Van Zon {\it et al.}
\cite{VanZon2007}. KaiC switches between an active conformation in
which the phosphorylation level tends to rise, and an inactive one in
which it tends to fall \cite{VanZon2007}. 
The model describes how KaiA binds to the CII domain of KaiC in the active conformation,
and how KaiA can then drive phosphorylation by acting as a nucleotide exchange factor
\cite{Nishiwaki-Ohkawa2014}, stimulating the exchange of ADP for ATP. The model
predicts that as the phosphorylation level in the CII domain rises,
the release of ADP from CI slows down. The ADP-bound state makes the inactive
conformation of KaiC more stable, causing the hexamer to flip to the
inactive state and triggering dephosphorylation. In
our model, ATP hydrolysis in the CI domain thus provides the
thermodynamic driving force for the oscillations, while the
phosphorylation in the CII domain provides the timer for the
hydrolysis in the CI domain.

While the coupling between ATP hydrolysis in the CI domain and
phosphorylation in the CII domain is the main feature of the new
model, leading to novel testable predictions, 
the model can also describe a wealth of additional experimental data. 
The differential affinity of KaiA for the
different KaiC phosphoforms naturally explains the characteristic
sequence in which the threonine and serine sites are phosphorylated
\cite{Nishiwaki2007,Rust2007}. 
In addition, while the slow release of ADP from CI triggers a switch between the active and inactive KaiC conformations, 
the binding of KaiB to CI stabilizes the inactive state further, 
and leads, as in previous models \cite{VanZon2007,Rust2007,Lin2014}, 
to the sequestration of KaiA, necessary for synchronizing the oscillations. 
The model predicts that the slow binding of KaiB, as observed experimentally \cite{Chang2015}, 
introduces a delay between the moment that a given KaiC hexamer reaches its point of maximum 
phosphorylation, and hence no longer needs KaiA to progress along
the phosphorylation cycle, and the moment that the same KaiC actually sequesters KaiA. 
In our model, this delay is essential because it allows the laggards 
(hexamers with a phosphorylation level lower than the mean) 
to reach the top of the cycle before the front runners 
(hexamers with a phosphorylation level higher than the mean)
take away KaiA. 
Lastly, the model can explain the experimental observation that the
oscillation period is robust to variations in steady-state ATP/ADP levels, 
while the system can be entrained by transient changes in this 
ratio, which is one of the important mechanisms for coupling the clock to light 
\cite{Pattanayak2015}.

\section*{Theory}

\subsection*{Model overview}
Our model of the in-vitro Kai circadian clock \cite{Nakajima2005} builds on the hexamer model
developed by Van Zon and coworkers \cite{VanZon2007}. 
But in contrast to that model, and following the models developed by
Rust {\em et al.} \cite{Rust2007,Phong2012,Lin2014}, it explicitly keeps track of the two phosphorylation
sites on each of the monomers, as well as their nucleotide binding states.  
The purpose of this section is to give an overview of the new model
and its state variables, and to provide background information on
ideas from previous models and their experimental justification. The
new ingredients of the model, as well as their experimental
movitation, are discussed only briefly; they are discussed in much
more detail in the sections below.

\textbf{KaiC Monomers} Our model follows the phosphorylation and
nucleotide-binding state of each of the six monomers inside a hexamer.  
Each monomer consists of two domains: The CI and the CII domain.  
The CII domain has two phosphorylation sites, 
the threonine and the serine site, 
resulting in four different phosphorylation states \cite{Rust2007, Nishiwaki2007}:
unphosphorylated (U), phosphorylated only on serine (S),
phosphorylated only on threonine (T), and phosphorylated on both
serine and threonine (D).  Furthermore, both the CI and CII domains
have a nucleotide binding pocket which can be in one of two possible
states: Either there is adenosine triphosphate (ATP) or adenosine
diphosphate (ADP) bound to it. The unbound state is ignored, because
nucleotide binding is much faster than nucleotide dissociation, as
described below. The state variables of the monomers
are given in \tref{StateVariables}.

In the next sections, we describe in detail how ATP hydrolysis in CI
drives the conformational switch of KaiC and how phosphorylation in CII
controls hydrolysis in CI.

\textbf{KaiC Hexamers} Van Zon {\it et al.} postulated that KaiC can be in
either an active (A) or inactive (I) conformation
\cite{VanZon2007}. Experiments probing the exposure of the C terminal
tails of KaiC and the stacking interactions between the CI and CII
domains indeed provide evidence that KaiC can exist in multiple conformational
states \cite{Kim2008,Chang2012,Egli2013,Snijder2014}. We follow Van
Zon {\it et al.}, and assume in the spirit of the Monod-Wyman-Changeux (MWC) model
\cite{Monod1965} that the CI and CII domains of all the
monomers in a hexamer switch conformation in concert, such that
we can speak of the hexamer as either being in the active or inactive state.
Following Van Zon {\it et al.} our model does not include monomer exchange,
which does not appear to be essential \cite{Rust2007,VanZon2007,Clodong2007}.

\textbf{KaiB binding} The phosphorylation behavior of KaiC in the
presence of KaiB, but not KaiA, is highly similar to that of KaiC
alone \cite{Xu2003,Kitayama2003, Rust2007}. This observation indicates that KaiB
does not directly affect the phosphorylation and dephosphorylation
rates. Following Van Zon {\it et al.} \cite{VanZon2007}, we assume instead
that KaiB plays the following dual role: $i)$ KaiB binding increases
the stability of the inactive state by binding to the CI domain of
inactive KaiC; experimental observations support the idea that
the binding of KaiB to KaiC depends on the conformational state of KaiC \cite{Chang2012,Egli2013,Snijder2014}; 
moreover, the experiments show that KaiB binding peaks in the
dephosphorylation phase of the cycle, when KaiC is in the inactive
conformational state \cite{Kageyama2003,Kageyama2006,Rust2007}; $ii)$
KaiB associated with the CI domain of inactive KaiC strongly
binds the limiting pool of KaiA, thereby sequestering it.  In our model, we
do not explicitly keep track of the KaiB concentration.  Recent
experiments show that KaiB binds KaiC as a monomer \cite{Snijder2014},
but is energetically most stable as a tetramer \cite{Chang2015}.
Because of the equilibrium between the terameric and monomeric state
of KaiB, the tetrameric state will act as a reservoir stabilizing the
KaiB monomer concentration, making the latter less dependent on the
total KaiB concentration \cite{Snijder2014}.  Moreover, as
  predicted by Van Zon {\it et al.} \cite{VanZon2007}, as long as KaiC can
bind enough KaiB to sequester KaiA effectively, the concentration of
KaiB does not affect the amplitude and period of the oscillations
\cite{Nakajima2010}. We therefore do not explicitly model the
concentration of KaiB, but rather include it in the definition of the
effective rate constant for KaiB-KaiC binding.

\textbf{KaiA binding} Experiments have unambiguously demonstrated that
KaiA stimulates the phosphorylation of KaiC \cite{Iwasaki2002, Kitayama2003, Xu2003, Nishiwaki2004}. 
Moreover, they indicate that in the absence of KaiB, KaiA binds to the CII domain \cite{Pattanayek2006}. 
Inspired by the recent observation that KaiA 
acts as a nucleotide exchange factor \cite{Nishiwaki-Ohkawa2014}, 
our new model describes how KaiA bound to CII is able to drive phosphorylation by
controlling the nucleotide exchange rate. In the presence of KaiB,
KaiA can also bind to the CI domain of inactive KaiC \cite{Chang2012}.
We do keep track of the KaiA dimers in the solution, and explicitly
model their binding to the CII domain and their sequestration on the CI
domain via KaiB.  The interactions are always described as
bimolecular reactions between KaiC hexamers and KaiA dimers or KaiB
monomers.  For simplicity, and lack of experimental evidence
suggesting otherwise, the binding of KaiA or KaiB to KaiC always
affects all monomers in the hexamer equally.  In our model, a single
KaiA dimer can bind to the CII domain of the hexamer, six KaiB
monomers can bind to the CI domain of a KaiC hexamer and six KaiA
dimers can, in turn, be sequestered by the CI domains of the hexamer
in the inactive state.  

{\textbf{State variables and parameters} The state variables
  describing the hexamer and possible values are summarized in
  \tref{StateVariables}. The parameters of the model were obtained by
  fitting the predictions of the model to experimental data, as
  explained in the sections below. In this procedure, the parameters
  were ``hand tuned''---we did not follow a systematic, formal,
  fitting procedure.
\begin{table}[hb]
\begin{center}
    \begin{tabular}{| l | c | c | l | c |} \cline{1-2} \cline{4-5}
    \multicolumn{2}{|c|}{\bf Monomer} & & \multicolumn{2}{c|}{\bf Hexamer} \\ \hhline{|-|-|~|-|-|} 
     \cellcolor{lightgray} Variable & \cellcolor{lightgray} States & & \cellcolor{lightgray} Variable & \cellcolor{lightgray} States \\ \hhline{|-|-|~|-|-|} 
     Phosphorylation & {U,T,D,S}  & & Conformation & Active/Inactive \\
     CI Binding pocket & ATP,ADP  & & \rmlabels{n}{CII\cdot KaiA}{} & $0,1$ \\ 
     CII Binding pocket & ATP,ADP & & \rmlabels{n}{CI\cdot KaiB}{} & $0-6$ \\ 
      &                           & & \rmlabels{n}{CI\cdot KaiA}{} & $0-6$ \\ \cline{1-2} \cline{4-5}
    \end{tabular}
\end{center}
\caption{\tlabel{StateVariables} Monomer and hexamer state variables, with possible values. 
Variables \rmlabels{n}{CI\cdot KaiA}{} and \rmlabels{n}{CII\cdot KaiA}{} 
count the number of KaiA dimers bound to the CI and CII domain, respectively, 
and \rmlabels{n}{CI\cdot KaiB}{} counts the number of KaiB monomers bound to CI.}
\end{table}

{\bf Article overview.} We have split the explanation of the full model into two parts: 
First we give a detailed description of the phosphorylation dynamics in the Kai system 
which primarily concerns the CII domain of KaiC and its interaction with KaiA.
In the next part, we describe the power cycle in the CI domain, 
the connection between the CI and CII domain 
and the binding and unbinding kinetics of KaiB and the subsequent sequestration of KaiA by the CI domain.
Then we present the results for the model, again split into two sections:
One relating to the phosphorylation dynamics and one to the power cycle in the CI domain.

\subsection*{Model of the KaiC phosphorylation dynamics}
Here we give a detailed description of how the phosphotranfer reactions, 
the ratio of ATP to ADP in the binding pockets and differential affinity of KaiA 
together give rise to the ordered phosphorylation of the serine and threonine sites in the CII domain.
In steps, we present the foundations of our model together with the experimental results 
that underlie it and give a detailed mathematical description 
of the resulting free energies and reaction rates.

\subsubsection*{Phosphorylation and dephosphorylation only occur via phosphotransfer with ATP and ADP}
Recent experiments unexpectedly showed that during the dephosphorylation of KaiC,
the inorganic phosphate group on the serine and threonine sites of KaiC is transfered to the ADP 
in the binding pocket of the CII domain,
effectively regenerating the ATP that was used for phosphorylation \cite{Nishiwaki2012, Egli2012}.
We hypothesize that in our model, dephosphorylation without a nucleotide as an intermediate does not occur.
Therefore, the phosphotransfer reactions are the only pathways for phosphorylation and dephosphorylation of KaiC,
\begin{eqnarray}
 \mathrm{U}\cdot \mathrm{ATP} \rates{\rmlabels{k}{0}{UT}}{\rmlabels{k}{0}{TU}} \mathrm{T}\cdot \mathrm{ADP}, \quad &
 \mathrm{T}\cdot \mathrm{ATP} \rates{\rmlabels{k}{0}{TD}}{\rmlabels{k}{0}{DT}} \mathrm{D}\cdot \mathrm{ADP}, \nonumber \\
 \mathrm{U}\cdot \mathrm{ATP} \rates{\rmlabels{k}{0}{US}}{\rmlabels{k}{0}{SU}} \mathrm{S}\cdot \mathrm{ADP}, \quad &
 \mathrm{S}\cdot \mathrm{ATP} \rates{\rmlabels{k}{0}{SD}}{\rmlabels{k}{0}{DS}} \mathrm{D}\cdot \mathrm{ADP}. 
 \elabel{CII_phosphotransfer}
\end{eqnarray}
Here, U,T,D and S correspond to the phosphorylation state of the monomer 
and ATP and ADP denote the state of the CII nucleotide binding pocket. 
\rmlabels{k}{0}{XY} are the phosphotransfer rate constants when KaiA is not bound to CII.
Since these rates are independent of the state of the other monomers,
the monomers in a hexamer are phosphorylated in a random order.
As is clear from \eref{CII_phosphotransfer}, 
the phosphorylation dynamics critically depends on the state of the nucleotide binding pocket of the CII domain: 
With ATP in the binding pocket, 
KaiC can only be phosphorylated and with ADP in the binding pocket KaiC can only be dephosphorylated.
Therefore, we explicitly keep track of the state of the nucleotide binding pocket adjacent 
to the serine and threonine sites of each monomer.

\subsubsection*{KaiA acts as a nucleotide exchange factor on the binding pockets of the CII domain}
{\bf Nucleotide exchange rate in absence of KaiA.} Since the nucleotide binding rates are much
faster than the dissociation rates, the unbound state can be neglected
\cite{Hayashi2003,Hayashi2004,Nishiwaki2012}, and the nucleotide
binding pocket will alternate only between the ATP and ADP bound
state, both on CI and CII.  Assuming that the association rates
for ATP and ADP are diffusion limited and similar, the dynamics of
nucleotide exchange will solely be governed by the nucleotide
dissociation rates, which for the CII domain, discussed here, are
denoted by \rmlabels{k}{CII\cdot ATP}{off} and
\rmlabels{k}{CII\cdot ADP}{off}, respectively.  Next to nucleotide exchange, ATP can
also be converted to ADP via hydrolysis with a rate
\rmlabels{k}{CII}{hyd} \cite{Terauchi2007,Nishiwaki2012}.  Since
hydrolysis is a strong downhill reaction under experimental
conditions, we neglect the reverse reaction of the hydrolysis pathway,
such that there is no ATP production by spontaneous binding of a
phosphate group to ADP.  The relative affinity between ATP and ADP for
the nucleotide binding pocket of the CII domain is given by
$\rmlabels{K}{CII}{ATP/ADP}=\rmlabels{K}{CII\cdot
  ATP}{d}/\rmlabels{K}{CII\cdot ADP}{d}=\rmlabels{k}{CII\cdot
  ATP}{off}/\rmlabels{k}{CII\cdot ADP}{off}$.  Assuming that the
nucleotide exchange and hydrolysis pathways are independent, we can
simply add their reaction rate constants, such that the rates for
changing between the ATP and ADP bound states become
\begin{eqnarray}
 \rmlabels{k}{CII}{ATP\to ADP} & = & \rmlabels{k}{CII}{hyd} + (1-\rmlabels{\alpha}{}{ATP})\,\rmlabels{k}{CII\cdot ATP}{off}, \elabel{CII_NucleotideExchange_ATPtoADP}\\
 \rmlabels{k}{CII}{ADP\to ATP} & = & \rmlabels{\alpha}{}{ATP}\,\rmlabels{k}{CII\cdot ADP}{off}.
 \elabel{CII_NucleotideExchange_ADPtoATP}
\end{eqnarray}
Here, \rmlabels{\alpha}{}{ATP} is the fraction, [ATP]/([ATP] + [ADP]), of ATP nucleotides in the solution.
The rate from ATP to ADP (\eref{CII_NucleotideExchange_ATPtoADP}) is the sum of the hydrolysis rate plus the rate of dissociating ATP times
the probability of immediately binding an ADP, which due to the equal
association rates for ADP and ATP binding,
is simply given by the fraction of ADP in the bulk $(1-\rmlabels{\alpha}{}{ATP})$.
The reverse rate (\eref{CII_NucleotideExchange_ADPtoATP}) is given by the rate of dissociating ADP 
times the probability, \ATPfrac, of binding an ATP nucleotide.

{\bf KaiA speeds up nucleotide exchange on CII domain.} 
It is well known that KaiA stimulates the phosphorylation of KaiC \cite{Iwasaki2002, Nishiwaki2004},
and that without KaiA, KaiC dephosphorylates \cite{Kitayama2003, Xu2003}.
Recent experiments showed that KaiA increases the fraction of ATP in
the nucleotide binding pockets and thereby stimulates phosphorylation \cite{Kitayama2013, Nishiwaki-Ohkawa2014}.
Without KaiA bound to the CII domain, the exchange rates between the nucleotide binding pocket 
and the bulk are very low,
such that eventually all ATP molecules bound to the binding pockets of the CII domain are hydrolyzed \cite{Nishiwaki2012, Nishiwaki-Ohkawa2014}.
Given these observations, in our model KaiA will act as a nucleotide exchange factor increasing the 
dissociation rates of both ATP and ADP in equal amounts, 
such that the relative affinity, $\rmlabels{K}{CII}{ATP/ADP}$, is unchanged.

We model the interaction between KaiA and the CII domain on three simplifying assumptions:
First, only one KaiA dimer can bind to the CII domain of a hexamer, 
although a higher stoichiometry has been observed \cite{Hayashi2004b, Brettschneider2010}.
Second, when KaiA is bound to CII, it enhances the nucleotide exchange rates in all the monomer binding pockets equally.
Third, when KaiA is not bound, the nucleotide dissociation rates are zero.
Therefore, the nucleotide dissociation rates given in \erefstwo{CII_NucleotideExchange_ATPtoADP}{CII_NucleotideExchange_ADPtoATP},
are given by \rmlabels{k}{CII\cdot ADP}{off, KaiA} \rmlabels{K}{CII}{ATP/ADP} 
and \rmlabels{k}{CII\cdot ADP}{off, KaiA}, respectively, when KaiA is bound to CI,
and equal zero when KaiA is absent.
The hydrolysis rate, $\rmlabels{k}{CII}{hyd}$, does not depend on whether KaiA is bound 
to the CII domain, in the interest of simplicity.

{\bf KaiA stimulates phosphorylation by speeding up nucleotide exchange.} 
\fref{CII_NucleotideExchange_ATPADP} illustrates how KaiA can enhance the ATP fraction 
in the nucleotide binding pocket of the CII domain. 
KaiA does not change the affinity of CII for ATP and ADP, 
and hence also leaves their relative affinity unchanged.
However, it does increase the binding and unbinding rates with the same magnitude. 
Moreover, the binding of the nucleotides is coupled to the non-equilibrium process of ATP hydrolysis,
which breaks detailed balance. 
The result is that in the absence of KaiA, 
the hydrolysis rate dominates over the nucleotide exchange rates,
driving the binding pockets towards the ADP state.
In the presence of KaiA, exchange rates are larger than the hydrolysis rate,
and because these rates favor ATP over ADP, the binding pocket is predominantly bound to ATP when KaiA is present.
By increasing the occupation of ATP of the binding pocket, 
KaiA not only enhances phosphorylation but also blocks dephosphorylation since there is no ADP. 
Blocking dephosphorylation, which was implicitly present in previous models \cite{VanZon2007, Rust2007, Lin2014}, 
is important, as it prevents futile phosphorylation cycles.

\begin{figure}[ht!]
\includegraphics[scale=1.0]{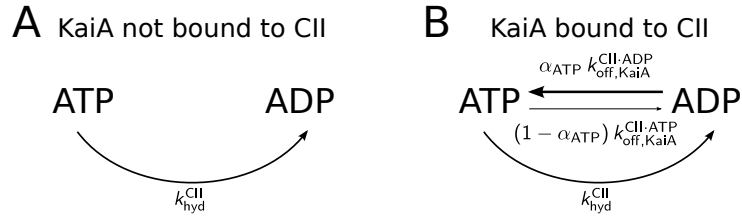} 
\caption{\flabel{CII_NucleotideExchange_ATPADP} KaiA regulates the
  fraction of ATP in the CII binding pockets by increasing the
  nucleotide dissociation rates, in combination with the ATPase
  activity in the CII domain.  A) Without KaiA bound to CII, the
  nucleotide dissociation rates are identically zero, and there is no
  nucleotide exchange with the bulk.  Since ATP is hydrolyzed at a
  constant rate, \rmlabels{k}{CII}{hyd}, eventually all the binding
  pockets will be occupied by ADP.  B) When KaiA is bound to the CII
  domain, it increases the dissociation rates of ADP and ATP,
  \rmlabels{k}{CII\cdot ATP}{off} and \rmlabels{k}{CII\cdot ADP}{off},
  respectively, while leaving the affinities for ATP and ADP
  unchanged.  Now, ADP in the CII domain is replaced by ATP at a rate
  that is faster than that at which ATP is hydrolyzed; this indeed
  increases the fraction of ATP in the binding pockets.  For
  simplicity, we assumed equal diffusion limited association rates for
  ATP and ADP, such that the probability of binding ATP after ADP has
  dissociated is equal to \rmlabels{\alpha}{}{ATP}.  The rate of the
  reverse pathway is proportional to the bulk ADP fraction,
  $1-\rmlabels{\alpha}{}{ATP}$.}
\end{figure}

\subsubsection*{Differential affinity: The affinity of KaiA for KaiC depends on phosphorylation state of KaiC.}
It was predicted from theoretical arguments \cite{VanZon2007}, 
and later confirmed by experiments \cite{Qin2010, Ma2012, Lin2014},
that the affinity of KaiC for KaiA depends on the phosphorylation state of the hexamer.
By measuring the phosphorylation speed of KaiC, starting with different initial phosphorylation levels,
Rust {\it et. al.} observed that the rate of KaiC phosphorylation decreases as the fraction of S and D phosphorylated 
KaiC monomers increases \cite{Lin2014}.
This suggests that KaiA has a high affinity when KaiC is in the U and T state, 
and a low affinity when KaiC is in the S and D state, 
leading to the mechanism of differential affinity \cite{VanZon2007}.

Differential affinity means that the binding and unbinding rates of
KaiA to the CII domain of KaiC depend on the phosphorylation state of
the hexamer. The observation that KaiA predominantly binds to CII
during the phosphorylation phase of the cycle indicates that, 
in addition, KaiA has a higher affinity for the active conformational state. 
In our model, when all the monomers of an active hexamer are
in the unphosphorylated U state,  KaiA will bind and unbind with the rates
$\rmlabels{k}{CII\cdot KaiA}{on,0}$ and $\rmlabels{k}{CII\cdot
  KaiA}{off,0}$, respectively.  The subsequent phosphorylation of KaiC
changes the binding free energy $\Delta\rmlabels{G}{CII\cdot
  KaiA}{bind}$: this indeed underlies the mechanism of differential
affinity.  Assuming each monomer adds linearly to
$\Delta\rmlabels{G}{CII\cdot KaiA}{bind}$, the change in the binding
free energy between KaiA and CII becomes
\begin{equation}
 \Delta\rmlabels{G}{CII\cdot KaiA}{bind} = 
 \sum_{i=1}^{6} \delta g^{\rm CII\cdot KaiA}_{\rm bind}(X_i) + 
 \rmlabels{h}{}{Inactive}\,\delta\rmlabels{g}{CII\cdot KaiA}{A,I}.
 \elabel{CII_DGKaiA}
%-\log\left(\frac{\rmlabels{k}{CII\cdot KaiA}{on,0}}{\rmlabels{k}{CII\cdot KaiA}{off,0}}\right)
 \end{equation}
Here, $\delta\rmlabels{g}{CII\cdot KaiA}{bind}(X_i)$ is the contribution of each monomer in phosphorylation
state $X_i\in\{\mathrm{U,T,D,S}\}$ to the binding free-energy.
KaiA bound to the CII domain stabilizes the active conformational state with a 
fixed free-energy difference $\delta\rmlabels{g}{CII\cdot KaiA}{A,I}$ and 
$\rmlabels{h}{}{Inactive}$ is an indicator function 
that is one when the hexamer is in the inactive state and zero otherwise.
Note that the stabilization of the active state with respect to the
inactive one does not depend on the phosphorylation state,
and therefore the hexamer's conformation will not affect the phosphotransfer dynamics \cite{Rust2007, Phong2012}.
Given the effect of differential affinity on the binding free energy,
$\Delta\rmlabels{G}{CII\cdot KaiA}{bind}$, detailed balance dictates that the association and dissociation rates
of KaiA become, respectively,
\begin{eqnarray}
\rmlabels{k}{CII\cdot KaiA}{on}  &=& \rmlabels{k}{CII\cdot
  KaiA}{on,0}  {\rm exp}\left( -(1-\lambda)\,\Delta
  \rmlabels{G}{CII\cdot KaiA}{\mathrm{bind}} \right) \\
\rmlabels{k}{CII\cdot KaiA}{off} &=& \rmlabels{k}{CII\cdot
  KaiA}{off,0} {\rm exp}\left( \lambda\,\Delta\rmlabels{G}{CII\cdot
    KaiA}{\mathrm{bind}} \right).
\end{eqnarray}
Assuming changes in binding free-energy have an equal effect on the association and dissociation rate, $\lambda=1/2$.

\subsubsection*{KaiA binding changes the phosphortansfer rates.} 
We model phosphorylation and dephosphorylation, with and without KaiA, 
via a microscopically reversible phosphotranfer reaction.  Then, as an
inevitable consequence of differential affinity, detailed balance
implies that the binding of KaiA must also influence the
phosphotransfer rates \cite{Boltzmann1872}. We denote the free-energy
difference between phosphorylation states X and Y, with X,Y$\in\{{\rm
  U,T,D,S}\}$, when KaiA is not bound to KaiC as
$\delta\rmlabels{g}{0}{XY}$ and when KaiA is bound as
$\delta\rmlabels{g}{KaiA}{XY}$. Detailed balance then implies that the
difference between these two, $\delta\rmlabels{g}{0}{XY} -
\delta\rmlabels{g}{KaiA}{XY}$, is equal to the change in the binding
free-energy of KaiA that results from a change in the phosphorylation
state ${\rm X}_i$ to ${\rm Y}_i$ of monomer $i$:
\begin{eqnarray}
\delta\rmlabels{g}{0}{XY} - \delta\rmlabels{g}{KaiA}{XY} & = & 
\Delta\rmlabels{G}{CII\cdot KaiA}{bind}({\rm Y}_i) - \Delta\rmlabels{G}{CII\cdot KaiA}{bind}({\rm X}_i) \elabel{ddg}\\
& = & \delta\rmlabels{g}{CII\cdot KaiA}{bind}({\rm Y}) - \delta\rmlabels{g}{CII\cdot KaiA}{bind}({\rm X}),
\end{eqnarray}
where $\delta\rmlabels{g}{0}{XY} =
-\log\left(\rmlabels{k}{0}{XY}/\rmlabels{k}{0}{YX}\right)$, with
\rmlabels{k}{0}{XY} given by \eref{CII_phosphotransfer}, and
$\delta\rmlabels{g}{KaiA}{XY} =
-\log\left(\rmlabels{k}{KaiA}{XY}/\rmlabels{k}{KaiA}{YX}\right)$.  The
second line follows after substituting \eref{CII_DGKaiA} for
$\Delta\rmlabels{G}{CII\cdot KaiA}{bind}$.  The subscript $i$ in
\eref{ddg} is to emphasize that we compare hexamers that differ in the
phosphorylation state of one of their monomers.  Finally, we can write
the phosphotransfer rate constants between states X and Y, for the
situation where KaiA is bound to CII, as
\begin{equation}
 \rmlabels{k}{KaiA}{XY} = \rmlabels{k}{0}{XY}\,\exp\left(-\frac{1}{2}\left[ \delta\rmlabels{g}{CII\cdot KaiA}{bind}({\rm Y}) - 
 \delta\rmlabels{g}{CII\cdot KaiA}{bind}({\rm X}) \right] \right),
\end{equation}
and the reverse reaction rate by interchanging labels X and Y.
This equation indeed shows that the phosphotranfer rates in the presence of KaiA, \rmlabels{k}{KaiA}{XY}, 
depend on how the phosphorylation states change the affinity for KaiA and the energy levels 
of the monomers, depicted in \fref{CII_EnergyLandscape}.

\begin{figure}[ht!]
\begin{center}
\includegraphics[scale=1.0]{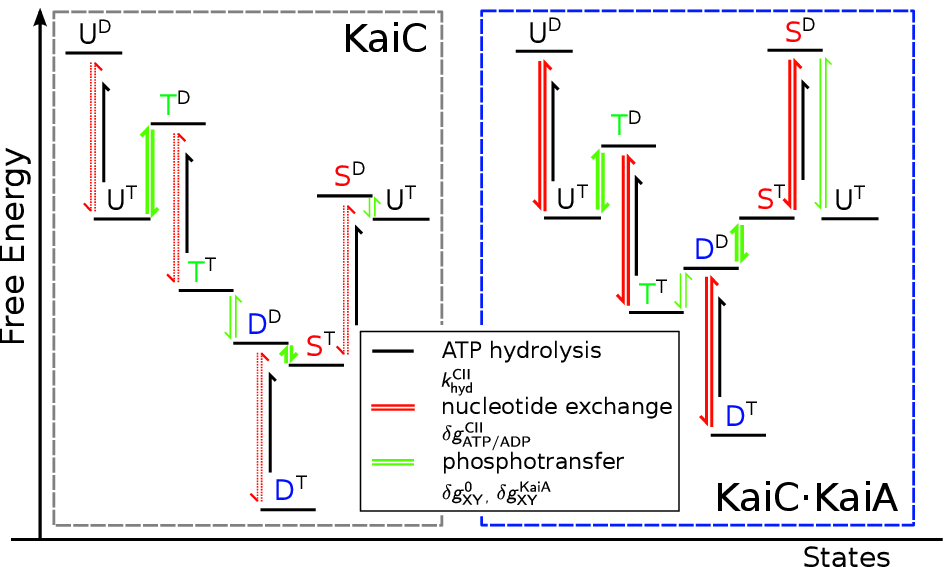} 
\end{center}
\caption{\flabel{CII_EnergyLandscape} Proposed free energy landscape of the CII domain, 
  without (left gray box) and with (right blue box) KaiA bound to the CII domain, 
  for equal concentrations of ATP and ADP in the bulk.
  Superscripts denote the nucleotide bound state of CII. 
  The landscape results from hand-tuning the
  phosphotransfer and nucleotide exchange parameters as well as the
  free-energy changes due to KaiA binding to give best agreement with
  the (de)phosphorylation assays performed in the SI of \cite{Phong2012}. 
  The free energy differences between phosphorylation states 
  are defined by the detailed-balance relations,
  $\delta\rmlabels{g}{0}{XY}=-\log\left(\rmlabels{k}{0}{XY}/\rmlabels{k}{0}{YX}\right)$ and
  $\delta\rmlabels{g}{KaiA}{XY}=-\log\left(\rmlabels{k}{KaiA}{XY}/\rmlabels{k}{KaiA}{YX}\right)$
  without and with KaiA bound to CII (connected by green arrows),
  respectively.  Thick and thin green arrows indicate high and low
  phosphotransfer rates, respectively.  Furthermore,
  $\delta\rmlabels{g}{CII}{ATP/ADP}=\log\left(\rmlabels{K}{CII}{ATP/ADP}\right)$,
  gives the free-energy difference between nucleotide bound states
  (connected by red arrows).  Note that the relative nucleotide
  affinity is independent of the KaiA-binding state or phosphorylation
  state.
  Dotted red arrows indicate the nucleotides exchange rates are identically zero 
  when KaiA is not bound to CII, 
  and solid red arrows in the right panel indicate high nucleotide exchange rates 
  when KaiA is bound.  
  The black arrows denote the irreversible hydrolysis of ATP.
  For hydrolysis, the magnitude of the arrow does not give the free energy change
  of the system since that involves the contribution from the phosphate release,
  which we do not take into account.
  Note that without hydrolysis, the total free-energy drop upon going through the cycle,
  $\rmlabels{U}{T}{}\rightarrow\rmlabels{T}{D}{}\rightarrow\rmlabels{T}{T}{}\rightarrow\rmlabels{D}{D}{}\rightarrow\rmlabels{S}{T}{}\rightarrow\rmlabels{S}{D}{}\rightarrow\rmlabels{U}{T}{}$,
  is zero.  }
\end{figure}

\subsubsection*{Ordered phosphorylation of the S and T sites due to phosphotransfer dynamics, nucleotide exchange and differential affinity}
It is well known that the S and T sites in each monomer are sequentially phosphorylated through the cycle: 
$\mathrm{U}\rightarrow\mathrm{T}\rightarrow\mathrm{D}\rightarrow\mathrm{S}\rightarrow\mathrm{U}$ \cite{Rust2007,Nishiwaki2007}.
In previous models, this order was imposed by choosing different effective rate constants between the 
phosphorylation states, in situations of KaiA bound to KaiC, and KaiA not bound to KaiC \cite{VanZon2007, Rust2007, Lin2014}.
However, given the properties of the phosphotransfer reactions and the effects of KaiA as described above,
we can now have a more detailed understanding of what makes the phosphorylation cycle go round.

At the start of the phosphorylation phase of the oscillation, 
the majority of monomers are in the non-phosphorylated state (U), 
with an ADP in the nucleotide binding pocket and a KaiA bound to the CII domain. 
The binding of KaiA will enhance the nucleotide exchange rate which increases the fraction of ATP 
in the binding pocket and consequently forces the phosphotransfer reactions in \eref{CII_phosphotransfer} towards 
the phosphorylated state while in the mean time blocking the reverse reaction.
The next question is why the threonine residue is phosphorylated before the serine residue.
Here, differential affinity plays a key role because KaiA binding lowers the free-energy
of the T-state and increases that of the S phosphorylation state, as is shown in \fref{CII_EnergyLandscape}.
Together with the fact that the phosphotransfer rate constants of the T site 
are much faster than those of the S site \cite{Rust2007, Egli2012}, 
the threonine residues are phosphorylated before the serine residues. 
Differential affinity thus has two important effects: 
Not only does it help to synchronize the KaiC hexamers as found in \cite{VanZon2007},
but it also enforces the correct order of phosphorylation.
Since a T-phosphorylated hexamer will still have a high affinity for KaiA, the ATP fraction 
in the binding pockets will remain high such that eventually both the serine and threonine sites are phosphorylated 
and the monomers arrive in the D state. 

In the dephosphorylation phase of the oscillation, 
when all KaiA is sequestered and therefore no nucleotide exchange is possible in the CII domain, 
the ATP in the CII binding pockets will eventually be hydrolyzed. 
With ADP in the binding pocket, phosphotransfer reactions can occur causing dephosphorylation \cite{Nishiwaki2012}. 
Without KaiA bound to CII, the serine residue becomes energetically favorable over the 
threonine residue again and because phosphotransfer with the threonine residue is faster than with the serine residue,
(meaning that the D$\rates{}{}$S transitions are faster than the D$\rates{}{}$T transitions)
the majority of the D phosphorylated monomers will proceed to the S state instead of the T state.
The S-site will slowly further dephosphorylate to the U state.
This shows how differential affinity and nucleotide exchange together give rise to the ordered
phosphorylation of the monomers.

\subsubsection*{ADP in solution slows down phosphorylation}
Experiments show that the fraction of ATP in solution, \linebreak $\rmlabels{\alpha}{}{ATP}=$[ATP]/([ATP]+[ADP]), 
has a significant effect on the phosphorylation speed and the
amplitude of the oscillations in the in-vitro system \cite{Rust2011,
  Phong2012, Pattanayak2015}. They also show that this sensitivity to the ATP fraction is the primary input for entraining the 
oscillator to the daily day-night cycle \cite{Rust2011, Phong2012, Pattanayak2015}.
As explained above, our model exhibits this sensitivity because the binding probabilities 
for ATP and ADP to the binding pocket of the CII domain are directly proportional to the ATP and ADP fraction, 
respectively, as given in \eref{CII_NucleotideExchange_ATPtoADP} and \eref{CII_NucleotideExchange_ADPtoATP}.

\subsection*{Model of the KaiC power cycle} 
In the previus section we discussed the phosphorylation dynamics of KaiC 
and the interaction between its CII domain and KaiA, 
and how these effects combine to generate the ordered phosphorylation of the threonine and serine sites in KaiC.  
The CI domain does not seem to play a crucial role here, in particular since
the phosphorylation dynamics in the presence of KaiA only, is
unaffected in a KaiC mutant where hydrolysis in the CI domain is
deactivated \cite{Phong2012}.  This raises the question of what role
the CI domain fulfills in the Kai oscillator.  Here we describe how
hydrolysis in CI together with the binding of KaiB to the CI domain
drives the conformational switch of the hexamer and how the slow
binding of KaiB, together with the subsequent sequestration of KaiA,
helps to synchronize the ensemble of KaiC hexamers.

\subsubsection*{Nucleotide exchange in and KaiB binding to the CI domain 
drives the conformational switch of KaiC}
{\bf KaiB and ADP binding to CI is cooperative.} 
Experiments show that the binding of KaiB requires catalytic
activity of the CI domain, since a mutant that lacks the hydrolysis
site does not bind KaiB \cite{Phong2012}.  Furthermore, when KaiB is
added to a solution with only KaiC and ATP, the ATPase rate drops
significantly \cite{Terauchi2007}. 
Because KaiA is not present in both experiments, 
the ATPase activity in the CII domain is negligible as explained 
in the section on KaiA acting as a nucleotide exchange factor, 
such that the change in the ATPase rate can be attributed to changes in the CI
domain.  Given these results, it seems likely that the affinity of
KaiB for KaiC depends on ADP in the CI binding pockets created by ATP
hydrolysis, and that, vice versa, KaiB binding stabilizes the binding
of ADP \cite{Snijder2014}.

In our model, the conformational switch from the active to inactive
state depends on ATP hydrolysis in the CI domain and the binding
of KaiB to the CI domain.  Specifically, both KaiB and ADP binding to
CI stabilize the inactive conformational state.  It is a
characteristic of the MWC model that this introduces an effective
cooperativity between KaiB and ADP: KaiB binding enhances the
probability that KaiC is in the inactive state, in which ADP will then
remain bound more strongly; conversely, ADP in CI will increase the
likelihood that KaiC is in the inactive state, in which it will bind
KaiB more strongly.

{\bf KaiB and ADP binding to CI stabilize the inactive conformational state.}
For simplicity, there is no direct cooperativity between ADP and KaiB
binding, such that the free-energy difference between the active and
inactive conformation of the hexamer is proportional to the number of
ADP nucleotides, \rmlabels{n}{CI\cdot ADP}{}, and KaiB monomers,
\rmlabels{n}{CI\cdot KaiB}{},
\begin{equation}
 \Delta\rmlabels{G}{hex}{A,I} = \left( \rmlabels{n}{CI\cdot ADP}{} - \rmlabels{n}{CI\cdot ADP}{0} \right)\,\delta\rmlabels{g}{ATP,ADP}{A,I} 
    + \rmlabels{n}{CI\cdot KaiB}{}\,\delta\rmlabels{g}{CI\cdot KaiB}{A,I}.
 \elabel{CI_DGATPADP}
\end{equation}
Here, $\delta\rmlabels{g}{ATP,ADP}{A,I} =
\delta\rmlabels{g}{ATP,ADP}{I} - \delta\rmlabels{g}{ATP,ADP}{A}$, is
the difference in the free-energy increase upon converting one ATP
into an ADP in the CI domain, between the inactive and active
conformational state of the hexamer.  Since the experiments indicate
that the stability of the inactive state increases with the number of
bound ADP molecules as discussed above, ADP needs to have a higher
affinity for the CI domain in the inactive state as compared to the
active state. On the other hand, the exchange of ADP for ATP should be
energetically favorable, $\delta\rmlabels{g}{ATP,ADP}{I} > 0$, such
that ADP can be exchanged spontaneously at the end of the
phosphorylation cycle. These two conditions can be satisfied by
choosing $\delta\rmlabels{g}{ATP,ADP}{A} >
\delta\rmlabels{g}{ATP,ADP}{I} > 0$ such that
$\delta\rmlabels{g}{ATP,ADP}{A,I} < 0$.  These conditions indeed
ensure that ATP hydrolysis stabilizes the inactive state, while still
allowing for spontaneous ADP release at the end of the cycle.  The
free-energy difference between the active and inactive state when all
CI binding pockets have ATP bound, is set in \eref{CI_DGATPADP} via
the parameter $\rmlabels{n}{CI\cdot ADP}{0}$, and determines the
threshold number of bound ADP molecules that are required to make the
inactive state more stable that the active one.  

Finally, the free-energy contribution to $\Delta\rmlabels{G}{hex}{A,I}$, from
binding a single KaiB monomer follows from the dissociation constants
for KaiB binding to the active and inactive state,
\rmlabels{K}{CI\cdot KaiB}{d,A} and \rmlabels{K}{CI\cdot KaiB}{d,I},
respectively, via the detailed balance relation \linebreak
$\delta\rmlabels{g}{CI\cdot KaiB}{A,I}=-\log\left(\rmlabels{K}{CI\cdot KaiB}{d,A}/\rmlabels{K}{CI\cdot KaiB}{d,I}\right)$. 
A higher affinity of KaiB for inactive KaiC than for active KaiC, 
$\rmlabels{K}{CI\cdot KaiB}{d,A}/\rmlabels{K}{CI\cdot KaiB}{d,I} > 1$, 
means that KaiB binding stabilizes the inactive state,
$\delta\rmlabels{g}{CI\cdot KaiB}{A,I} < 0$. 
\eref{CI_DGATPADP} thus shows how ADP and KaiB binding stabilize the inactive conformational state of KaiC.

\subsubsection*{Timing of the conformational switch is determined by
phosphorylation of CII domain, which sets ADP dissociation rate in CI domain}
The switch from the active to the inactive state is driven energetically 
by hydrolysis of ATP in the CI domain and the subsequent binding of KaiB.
But what exactly sets the timing of this switch?

{\bf Phosphorylation of the CII domain controls ADP dissociation rate in CI domain.}
Interestingly, experiments show that the binding of KaiB requires not
only the hydrolysis of ATP in CI, but also that the CII domain is
phosphorylated at least on the serine residue \cite{Phong2012}.  The
latter observation might be the result of a direct interaction of KaiB
with the CII domain, but could also be due to an indirect effect, in
which the likelihood that CI is bound to ADP (which enhances KaiB
binding), depends on the phosphorylation state of CII.  The latter
hypothesis is supported by the experimental observation that
hyperphosphorylated KaiC has a lower ATPase activity and a higher
fraction of ADP in the binding pockets, as compared to
non-phosphorylated KaiC \cite{Terauchi2007, Murakami2008,
  Nishiwaki-Ohkawa2014}.  As mentioned before, the lower measured
ATPase activity in hyperphosphorylated KaiC must, because of the
absence of KaiA in these experiments, be attributed to the CI domain,
and not to changes in the CII domain.  The lower ATPase rate is the
result of a lower hydrolysis rate and/or a lower ADP dissociation
rate.  However, a lower hydrolysis rate with a constant ADP
dissociation rate would lead to a lower ADP fraction in the binding
pockets, in contrast to what has been observed experimentally
\cite{Nishiwaki2012, Nishiwaki-Ohkawa2014}. 
We thus conclude that the phosphorylation state of CII
determines the ATPase rate of CI through the ADP dissociation rate: 
As serine residues on CII become phosphorylated, the ADP dissociation
rate at the CI domain decreases.  These arguments indicate that the
regulatory mechanism that controls the timing of the conformational
switch is the dependence of the dissociation rate of ADP in the CI
domain on the phosphorylation state of the CII domain.

{\bf The ADP-CI association and dissociation rates change, but their ratio, the affinity, does not.} 
A question not yet answered is whether the phosphorylation of the CII
domain changes the magnitude of the ADP binding rates, or also the
affinity, which depends on the ratio of the association and
dissociation rates.  Recent experiments allow us to answer this
question.  These experiments show that both the phosphorylation and
dephosphorylation rates are unchanged in a KaiC mutant where ATP
hydrolysis in the CI domain is deactivated, which decreases the
fraction of bound ADP \cite{Phong2012}.  
%Blocking hydrolysis in CI decreases the ADP fraction.  
Detailed balance would entail that if phosphorylation of CII were to stabilize ADP in CI,
then vice versa ADP would stabilize the phosphorylated state; 
KaiC hexamers with fewer ADP molecules in CI, 
such as the KaiC mutant in \cite{Phong2012}, would then dephosphorylate faster.
\fref{CICII_Connection}A illustrates how the detailed balance
condition changes the phosphotranfer rates in this case.
Since the experiments show that the (de)phosphorylation rates are unchanged \cite{Phong2012}, 
we must conclude that the affinity of ADP for CI does not depend on the
phosphorylation state of the CII domain.  Hence not only the
dissociation rate of ADP changes, but also the association rate
changes by the same factor, leaving the affinity unchanged.

{\bf Phosphorylation of CII controls ADP dissociation from CI via transition state.}
To explain the dependence of the ADP fraction in the CI domain on the phosphorylation state of the CII domain,
we envision a model in which the phosphorylation state of CII affects a short-lived transition state 
for the dissociation of ADP from the nucleotide binding pocket of CI. 
In our model, the activation energy for ADP dissociation in each monomer, $\Delta\rmlabels{G}{CI\cdot ADP}{act}$,
depends linearly on the phosphorylation state $X_i$ of all monomers in the hexamer
\begin{equation}
 \Delta\rmlabels{G}{CI\cdot ADP}{act}=\sum_{i=1}^{6} \delta\rmlabels{g}{CI\cdot ADP}{act, A/I}(X_i),
 \elabel{CI_dGact_ADPdissociation}
\end{equation}
where $\delta\rmlabels{g}{CI\cdot ADP}{act, A/I}$ is the contribution
of a single monomer on the activation energy in the active (A) or
inactive (I) conformational state. The ADP dissociation rate is then
given by
\begin{equation}
 \rmlabels{k}{CI\cdot ADP}{off} =  \rmlabels{k}{CI\cdot ADP}{off,
   0}\,\exp\left(-\Delta\rmlabels{G}{CI\cdot ADP}{act} \right),
\elabel{koffADPCI}
\end{equation}
where $\rmlabels{k}{CI\cdot ADP}{off, 0}$ is the off-rate when the hexamer is in the active state with 
all monomers in the U-state.

\fref{CICII_Connection}B shows the
ADP dissociation rate as a function of the number of monomers in the S
state for a hexamer in the inactive conformation, assuming that the
other\, mo\-nomers are in the U state, which is typically the case during
the dephosphorylation phase of the cycle.  The energy values
$\delta\rmlabels{g}{CI\cdot ADP}{act, I}$(X) determine when the rate
of ADP dissociation (typically leading to ATP binding) will be higher
than the ATP hydrolysis rate (leading to the ADP bound state). 
We choose the energy values such that the crossover in the rates happens
between 1 and 2 monomers in the S-state. Hence, when a hexamer has
fewer than 2 monomers in the S-state, ADP will be released and ATP
becomes bound, and the hexamer will switch back to the active
conformation, completing the cycle.

\begin{figure}[th!]
\begin{center}
\includegraphics[scale=0.8]{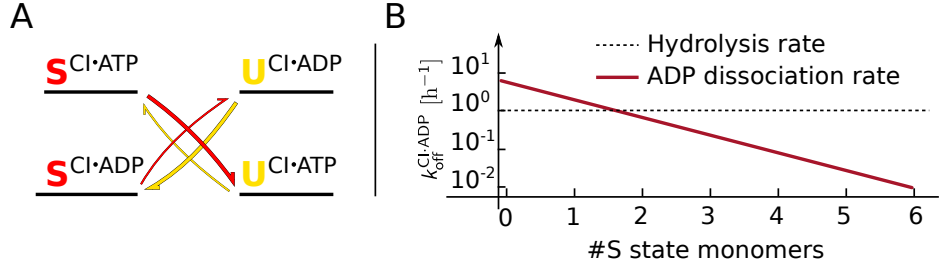} 
\end{center}
\caption{\flabel{CICII_Connection} The phosphorylation state of the
  CII domain regulates the ADP fraction in the CI domain by changing
  the ADP release rate, but not its affinity.  (A) Cartoon that
  illustrates why the experiments rule out the latter ``affinity''
  scenario. The cartoon shows a free-energy landscape for this
  affinity scenario in which the phosphorylation state of CII
  determines the affinity of ADP for CI; specifically, the cartoon
  illustrates the case where a monomer in the S state stabilizes ADP
  in the binding pocket of the CI domain.  The superscripts denote the
  state of the CI binding pocket.  The S-phosphorylated state lowers
  the free energy of the CI$\cdot$ADP state compared to the U state,
  thereby decreasing the ADP-off rate in the CI binding pocket.  It
  follows from detailed balance that when the CI domain has ADP bound,
  the phosphotransfer from S to U is uphill in free energy, which
  decreases the phosphotransfer rate.  The diagonal arrows give the
  phosphotransfer reactions from U to S (yellow) and S to U (red).
  When the CI domain is predominantly in the ATP state, as was done in
  \cite{Phong2012} by lowering the hydrolysis rate with a mutation,
  the affinity scenario, as illustrated in the cartoon, would predict
  that the dephosphorylation rate from S to U increases.  However, the
  experiments show no significant change in phosphotransfer rates,
  making this affinity scenario unlikely. Based on these arguments, we
  predict that phoshorylation of CII does not change the
  affinity, but does change the absolutes rates of ADP dissociation and
  association. (B) ADP dissociation rate as a function of the number
  of S state monomers in the hexamer, assuming the other monomers are
  in the U state. Between 1 and 2 monomers in the S state, the
  dissociation rate is higher than the hydrolysis rate which puts the
  binding pockets of CI domain predominantly in the ATP state, driving
  the hexamer to the active conformational state.}
\end{figure}

{\bf ADP-CI association rate is very low and ADP arises only via hydrolysis of bound ATP.} 
Because ADP in the CI binding pocket is energetically very unfavorable \cite{Mori2002,Hayashi2003,Hayashi2004}, 
and the affinity of KaiC for KaiB does not seem to depend on the bulk ATP fraction \cite{Rust2011,Phong2012}, 
the association rate of ADP from the bulk to the CI binding pocket will be very low.  
Therefore, in our model, regardless of the bulk ATP fraction, 
ADP can only appear in the CI binding pocket through the hydrolysis of ATP:
\begin{equation}
 \rmlabels{k}{CI}{ATP\to ADP}  =  \rmlabels{k}{CI}{hyd}.
\end{equation}
%Because the nucleotide dissociation rates are much lower than the association
%rates, the CI binding pockets are always occupied with either ATP or
%ADP \cite{Nishiwaki-Ohkawa2014}. 
Moreover, since the ADP association rate is assumed to be zero, after
ADP dissociation the pocket will always bind ATP. 
Since ATP association is much faster than nucleotide dissociation
\cite{Nishiwaki-Ohkawa2014}, as exploited in the section on phosphorylation dynamics, 
the rate of exchanging ADP for ATP is simply given by the ADP dissociation rate:
\begin{equation}
 \rmlabels{k}{CI}{ADP\to ATP} =   \rmlabels{k}{CI\cdot ADP}{off},
\end{equation}
with the latter given by \eref{koffADPCI}. 
%An important observation is that KaiB does not change the dephosphorylation rate of an ensemble of highly phosphorylated KaiC
% Indeed, experiments in \cite{Phong2012} show that the ATPase rate of a CI-only mutant of KaiC is almost independent
% of the ATP fraction in the bulk.
% Furthermore, KaiC monomers in solution with ADP will not form hexamers showing a very low affinity for ADP 
% \cite{Mori2002,Hayashi2003}.

\subsubsection*{T and S phosphorylation states have an antagonistic effect on the ADP fraction in CI} 
Experiments show that a phosphomimetic of the T state of KaiC has a higher ATPase activity
compared to unphosphorylated KaiC and that a phosphomimetic of the S state has a lowered ATPase activity \cite{Murayama2011}.
This is likely the result of an antagonistic effect of phosphorylation of the threonine and serine sites on the dissociation rate of ADP.
Specifically, in our model monomers in the T state will lower the activation energy for ADP release,
$\delta\rmlabels{g}{CI\cdot ADP}{act}(\rm{T})<0$, while monomers in the S state increase 
the activation energy, $\delta\rmlabels{g}{CI\cdot ADP}{act}(\rm{S})>0$. 
Because the S and T sites are orderly phosphorylated, 
their antagonistic effect on the dissociation rate of ADP will create a sharp transition between the phase in which 
the ADP fraction in CI is low and that in which it it is high \cite{Lin2014}.
Furthermore, just like in the push-pull network studied by Goldbeter
{\it et al.} \cite{Goldbeter1981},
the ATP fraction in the CI domain will depend on the \emph{difference} between the monomers in the S and T 
state, and not on their absolute number.
This makes the regulation of the CI domain less sensitive to the absolute phosphorylation level,
which depends on the bulk ATP fraction \cite{Rust2011, Phong2012}. 

\subsubsection*{Nucleotides, KaiB and KaiA binding to the CI domain stabilizes the inactive state}
Detailed balance implies that the different affinities of the binding partners for the active and inactive state of KaiC,
is reflected in the free-energy difference between the two conformations, $\Delta\rmlabels{G}{hex}{A,I}$.
In our model, we assume there is no cooperative binding to KaiC in a
given conformational state (although the MWC model introduces an
effective cooperativity as explained in the theory section on the power cycle),
such that we can split the dependence on each binding partner in independent terms,
\begin{eqnarray}
 \Delta\rmlabels{G}{hex}{A,I} & = & \left(\rmlabels{n}{CI\cdot ADP}{} - \rmlabels{n}{CI\cdot ADP}{0}\right)\,\delta\rmlabels{g}{ATP,ADP}{A,I} + \rmlabels{n}{CI\cdot KaiB}{}\,\delta\rmlabels{g}{CI\cdot KaiB}{A,I} \nonumber \\  
  & \phantom{ = } & + n^{\rm CI\cdot KaiA}\,\delta\rmlabels{g}{CI\cdot KaiA}{A,I} + n^{\rm CII.KaiA}\,\delta\rmlabels{g}{CII\cdot KaiA}{A,I}. 
 % \Delta\rmlabels{G}{ATP,ADP}{A,I} + \Delta\rmlabels{G}{CI\cdot KaiB}{A,I} + \Delta\rmlabels{G}{CI\cdot KaiA}{A,I} + \Delta\rmlabels{G}{CII\cdot KaiA}{A,I}\nonumber \\
 %  & = & \left(\rmlabels{n}{CI\cdot ADP}{} - \rmlabels{n}{CI\cdot ADP}{0}\right)\,\delta\rmlabels{g}{ATP,ADP}{A,I} + n^{\rm CI\cdot KaiB}\,{\rm log}\left( \frac{\tilde{K}^{\rm CI\cdot KaiB}_{\rm d}}{K^{\rm CI\cdot KaiB}_{\rm d}} \right) \nonumber \\  
 %  & \phantom{ = } & + n^{\rm CI\cdot KaiA}\,{\rm log}\left( \frac{\tilde{K}^{\rm CI\cdot KaiA}_{\rm d}}{K^{\rm CI\cdot KaiA}_{\rm d}} \right) + n^{\rm CII.KaiA}\,{\rm log}\left( \frac{\tilde{K}^{\rm CII\cdot KaiA}_{\rm d}}{K^{\rm CII\cdot KaiA}_{\rm d}} \right). 
  \elabel{DGAI}  
\end{eqnarray}
Each contribution is directly proportional to the the number of ADP nucleotides, KaiB monomers and KaiA dimers, 
\rmlabels{n}{CI\cdot ADP}{}, \rmlabels{n}{CI\cdot KaiB}{}, and \rmlabels{n}{CI\cdot KaiA}{},
respectively, bound to the CI domain.
The last contribution is proportional to the number of KaiA dimers, $n^{\rm CII\cdot KaiA}$, bound to the CII domain.
The terms depending on the number of ADP and KaiB proteins bound are explained in the theory section.
As discussed in more detail in the next section,
KaiA can only bind to the CI domain when 6 KaiB monomers are bound to
CI; moreover, 6 KaiA dimers can then be sequestered.
The free-energy contribution for binding a single KaiA dimer to CI results from the detailed-balance relation,\newline
$\delta\rmlabels{g}{CI\cdot KaiA}{A,I}=-\log\left({\rmlabels{K}{CI\cdot KaiA}{d,A}}/{\rmlabels{K}{CI\cdot KaiA}{d,I}}\right)$,
where \rmlabels{K}{CI\cdot KaiA}{d,A} and \rmlabels{K}{CI\cdot KaiA}{d,I} are the dissociation constants
for the active and inactive state of KaiC, respectively.
A similar relation holds for KaiA binding to CII. 

Because KaiA and KaiB have a higher affinity for inactive KaiC than for active KaiC, their binding to the CI domain will
stabilize the inactive state. This, together with CI-ATP stabilizing
the active state and CI-ADP stabilizing the inactive one, creates a
hysteresis loop, as illustrated in \fref{Energies_CI}. Importantly,
while KaiA and KaiB binding to CI stabilizes the inactive state with
respect to the active one, the system is designed such that when no
ADP is bound to CI, the active conformation with KaiA and KaiB bound
is {\em more} stable than the inactive one with KaiA and KaiB bound to
it. This is critical, because it ensures that when, during the
dephosphorylation phase, the system eventually reaches the point where
all ADP has been released and the number of CI-bound ADP molecules has reached
zero, the hexamer flips back from the inactive to the active state. In this
active state, KaiA and KaiB will then spontaneously dissociate from
KaiC, because this conformational state has a low affinity for KaiA
and KaiB.
Given \eref{DGAI}, the requirement that the active state with KaiA and KaiB bound 
is always more stable than the inactive one when no ADP is bound to CI, entails that 
$\rmlabels{n}{CI\cdot ADP}{0}\,\delta\rmlabels{g}{ATP,ADP}{A,I} \leq
\rmlabels{n}{CI\cdot KaiA}{max}\,\delta\rmlabels{g}{CI\cdot KaiA}{A,I}
+ \rmlabels{n}{CI\cdot KaiB}{max}\,\delta\rmlabels{g}{CI\cdot
  KaiB}{A,I}$.

\begin{figure}[ht!]
\begin{center}
\includegraphics[scale=1.0]{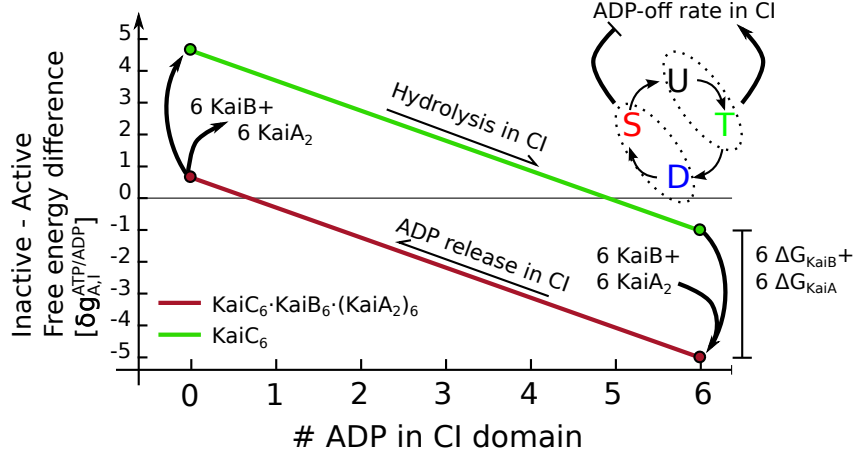} 
\end{center}
\caption{\flabel{Energies_CI} Hydrolysis of ATP in the CI domain, and
  the subsequent binding op KaiB and KaiA, forms a hysteresis loop in
  the free-energy difference between the active and inactive state.
  Starting in the active state with no ADP bound to CI, the
  free-energy difference linearly decreases as ATP is hydrolyzed to
  ADP, as indicated by the green line.  The number of ADP in CI is set
  by the competition between a fixed hydrolysis rate and a variable
  ADP off-rate, which is set by the phosphorylation state of the CII
  domains of the whole hexamer.  Phosphorylation of the T site
  initially enhances the ADP off-rate, but the subsequent
  phosphorylation of the S-site will decrease the dissociation of ADP.
  The antagonistic effect of the cyclically phosphorylated T and S
  sites on the ADP dissociation rate causes a sharp transition in the
  CI binding pocket from ATP dominated to ADP dominated \cite{Lin2014}. 
  When there are 5 or 6 ADP nucleotides in the CI domain, the hexamer will flip
  to the inactive state, increasing the affinity for KaiB, which will
  then slowly bind inactive KaiC.  When 6 KaiB monomers are bound to
  the hexamer, up to six free KaiA dimers will be sequestered from
  solution.  This complex of KaiB and KaiA on the CI domain stabilizes
  the inactive state of KaiC further (red line).  When all KaiA is
  sequestered, the dephosphorylation of the S-site in the CII domain
  will increase the nucleotide exchange rate in the CI domain,
  decreasing the ADP level and increasing the free-energy difference
  between the active and inactive state.  This allows the hexamer to
  flip back to the active state. In the active state, KaiA and KaiB
  dissociate from KaiC, because they have a low affinity for CI in the active
  conformation.}
\end{figure}
  
Given  $\Delta\rmlabels{G}{hex}{A,I}$, the rates of switching from the active to the inactive state, 
\rmlabels{k}{conf}{f}, and vice versa, \rmlabels{k}{conf}{b}, become
$\rmlabels{k}{conf}{f} = \rmlabels{k}{conf}{0} {\rm exp}\left(-\Delta\rmlabels{G}{hex}{A,I}/2\right)$
and $\rmlabels{k}{conf}{b} = \rmlabels{k}{conf}{0} {\rm exp}\left(\Delta\rmlabels{G}{hex}{A,I}/2\right)$,
respectively. The prefactor \rmlabels{k}{conf}{0} sets the timescale of switching.
%Combinatorics cancels in eq_DGAI since it is equal in both conformations.
% Here $\Delta\rmlabels{G}{ATP,ADP}{A,I}$ is the contribution by the nucleotides bound to CI, 
% $\Delta\rmlabels{G}{CI\cdot KaiB}{A,I}$ and $\Delta\rmlabels{G}{CI\cdot KaiA}{A,I}$ 
% are the contributions of KaiB and KaiA bound to the CI domain, 
% respectively, and $\Delta\rmlabels{G}{CII\cdot KaiA}{A,I}$ is due to KaiA bound to the CII domain.

\subsubsection*{Slow KaiB binding sets a time delay between the phosphorylation and dephosphorylation phase}
Experiments show the binding of KaiB to the KaiC hexamer is slow due
to a combination of a slow conformational switch of KaiB monomers in
solution, and a slow reaction step attributed to the CI domain of KaiC
\cite{Phong2012, Chang2015}.  Theoretical modeling suggests that the
Kai oscillator requires a time delay between the phase of phosphorylation
and the phase of KaiA sequestration and dephosphorylation, in order to generate
stable oscillations \cite{VanZon2007, Zwicker2010, Lin2014} and
a period that is fairly insensitive to changes in the bulk ATP fraction
\cite{Phong2012}.  We introduce this delay by assuming KaiC can not
sequester KaiA before a full ring of 6 KaiB monomers has formed on the
CI domain \cite{Villarreal2013, Snijder2014}, and that the rate of
KaiB binding to KaiC is slow and independent of the number of KaiB
proteins already bound.  In this way we simulate the slow appearance of
KaiB monomers in the bulk that have a binding competent conformation \cite{Chang2015}. 
As explained in the model overview, 
we do not explicitly keep track of KaiB, 
but coarse grain the KaiB concentration in the association rates motivated by the
observation that the absolute concentration of KaiB has little
influence on the amplitude and period of the oscillation
\cite{VanZon2007, Nakajima2010}.

After a KaiB ring has formed, KaiA will immediately be sequestered from solution
due to a very high on-rate, with a maximum of 6 KaiA dimers per hexamer \cite{Brettschneider2010, Qin2010}.
The affinity of KaiA for the hexamer with a KaiB ring depends on the conformational state of the hexamer:
Only in the inactive state KaiA stays bound to KaiC-KaiB. 
In the active state of KaiC, the CI domain has a lower affinity for
both KaiA and KaiB, as discussed in the previous section.
Hence, after KaiC has flipped to the active state, KaiA and KaiB will be released, and the cycle starts over.

\subsubsection*{Summary of the cycle dynamics}
Hydrolysis of ATP in the CI domain drives the conformational transition from the active state to the inactive one,
because ADP in the CI domain stabilizes the inactive state with respect to the active one, see \fref{Energies_CI}. 
This allows the ensemble of KaiC hexamers to switch from a phosphorylation phase with free KaiA in solution 
to a dephosphorylation phase where all KaiA is sequestered by KaiC.
In the inactive state, the CI domain of KaiC has a high affinity for both KaiA and KaiB,
meaning that the complex ${\rm KaiC_6 \cdot KaiB_6 \cdot}$ ${\rm (KaiA_2)_6}$, is energetically very stable.

While all KaiA is sequestered, the KaiC ensemble will dephosphorylate leaving most of the monomers in the U or S state.
When the number of S phosphorylated sites drops below a threshold, the
energy barrier for ADP release from CI will become sufficiently small,
such that ADP will dissociate even though ADP is
stabilized by the binding of KaiA and KaiB.
Without ADP bound to the CI domain the hexamer returns to the active state,
which has a lower affinity for KaiA and KaiB.
The sequestered proteins are then immediately released such that the cycle can start over again 
with an ensemble of hexamers with monomers in the U state.

% Place figure captions after the first paragraph in which they are cited. Add figures seperately.
\section*{Results}
\subsection*{Results on phosphorylation dynamics}
To test the validity of our model of the CII domain and to find the correct parameter values
shown in \tref{CII_Parameters},
we compare with the rich body of quantitative experimental results on the in-vitro Kai system.
As is done in these experiments, we will study the behavior of different combinations of the main actors: 
KaiA, KaiB and KaiC and the ATP fraction, \rmlabels{\alpha}{}{ATP}.
First, we will study the dephosphorylation dynamics of phosphorylated KaiC in the absence of KaiA and KaiB.
We will investigate the dependence on the ATP fraction and compare to experimental results.
Furthermore, we test our hypothesis that phosphotransfer from the threonine and serine sites to ADP 
is the only possible pathway for dephosphorylation, 
by comparing to a model where release of the phosphate into the bulk is possible.
Next, we study the effects of KaiA on the phosphorylation dynamics and the influence of
the bulk ATP fraction on the speed and steady state level of phosphorylation.
Lastly, we study if the ordered phosphorylation dynamics of the serine
(S) and threonine (T) residues persists when the system of KaiA and KaiC 
has reached steady state.
In the subsequent section, we address the role of KaiB and the oscillatory dynamics.
All simulations were performed at the experimental standard concentrations of 0.6$\upmu$M 
for KaiA and KaiC, which corresponds to simulating 720 KaiC hexamers and 720 KaiA dimers 
in a volume of 2 cubic micron.

\begin{table}[ht!]
\begin{center}
    \begin{tabular}{| l | c | c |} \hline
    \multicolumn{3}{|c|}{}  \\[-7pt]
    \multicolumn{3}{|c|}{ \bf \large{Parameters relating to the CII domain}}  \\[3pt] \hline \hline
     \rowcolor{lightgray} Parameter & Value & Explanation \\ \hline 
     \multicolumn{3}{|c|}{\bf Phosphotransfer} \\ \hline
     \rmlabels{k}{0}{UT} & 0.50 h$^{-1}$ & U$\cdot$ATP$\rightarrow$T$\cdot$ADP \\ 
     \rmlabels{k}{0}{TU} & 1.78 h$^{-1}$& T$\cdot$ADP$\rightarrow$U$\cdot$ATP \\ 
     \rmlabels{k}{0}{TD} & 0.40 h$^{-1}$& T$\cdot$ATP$\rightarrow$D$\cdot$ADP \\ 
     \rmlabels{k}{0}{DT} & 0.20 h$^{-1}$& D$\cdot$ADP$\rightarrow$T$\cdot$ATP \\ 
     \rmlabels{k}{0}{SD} & 1.50 h$^{-1}$& S$\cdot$ATP$\rightarrow$D$\cdot$ADP \\ 
     \rmlabels{k}{0}{DS} & 2.00 h$^{-1}$& D$\cdot$ADP$\rightarrow$S$\cdot$ATP \\ 
     \rmlabels{k}{0}{US} & 0.15 h$^{-1}$& U$\cdot$ATP$\rightarrow$S$\cdot$ADP \\ 
     \rmlabels{k}{0}{SU} & 0.20 h$^{-1}$& S$\cdot$ADP$\rightarrow$U$\cdot$ATP \\ \hline \hline
     \multicolumn{3}{|c|}{\bf Nucleotide binding pocket } \\ \hline
     %Parameter & Value & Explanation \\ \hline     
     \rmlabels{k}{CII}{hyd} & 1.00 h$^{-1}$ & ATP hydrolysis rate \\ 
     \rmlabels{k}{CII\cdot ADP}{off, KaiA} & 6.00 h$^{-1}$ & ADP off-rate with KaiA bound \\ 
     \rmlabels{K}{CII}{ATP/ADP} & 0.10 & Relative affinity for ATP and ADP \\ \hline \hline
     \multicolumn{3}{|c|}{\bf KaiA affinity } \\ \hline
     %Parameter & Value & Explanation \\ \hline 
     \rmlabels{k}{CII\cdot KaiA}{on, 0}  & \phantom{-}1.00 mM h$^{-1}$ & KaiA on-rate for CII domain \\ 
     \rmlabels{k}{CII\cdot KaiA}{off, 0} & \phantom{-}1.00 h$^{-1}$ &  KaiA off-rate for CII domain \\ 
     $\delta \rmlabels{g}{CII\cdot KaiA}{bind}(U)$ & 0.00 kT &  Free energy effect of U-monomer \\ 
     $\delta \rmlabels{g}{CII\cdot KaiA}{bind}(T)$ & -0.30 kT &  Free energy effect of T-monomer \\ 
     $\delta \rmlabels{g}{CII\cdot KaiA}{bind}(D)$ & 1.00 kT &  Free energy effect of D-monomer \\ 
     $\delta \rmlabels{g}{CII\cdot KaiA}{bind}(S)$ & 2.00 kT &  Free energy effect of S-monomer \\ 
     $\Delta \rmlabels{G}{CII\cdot KaiA}{A,I}$     & 10 kT & Free energy effect of conformation \\ \hline
     \end{tabular}
\end{center}
\caption{\tlabel{CII_Parameters} Model parameters relating to the CII domain
are introduced in the theory section on phosphorylation dynamics and their values 
are motivated in the results section.
Energies are given in units of kT, were k is Boltzmann's constant and T the temperature.
}
\end{table}

\subsubsection*{KaiC dephosphorylation speed is set by phosphotransfer rates}
{\bf Dephosphorylation via phosphotransfer can reproduce experiments.}
We start with simulating an ensemble of KaiC hexamers in a 100\% ATP
solution, and compare with the experimental results presented in the
supplementary information of \cite{Phong2012}.  Initially, KaiC is
highly phosphorylated.  As there is no KaiA (and also no KaiB), this
experiment allows us to distinguish the rate constants related to
phosphotransfer dynamics given in \eref{CII_phosphotransfer}, from the
effects related to the interaction with KaiA.  To obtain the rapid
decay of the T and D phosphorylated states and the transient peak in
the S state, as shown in experiments, the phosphotransfer rates
relating to the threonine site have to be significantly faster than
the rates relating to the serine site.  Furthermore, dephosphorylation
is downhill in free energy.  Otherwise, the ATP regenerated in the
dephosphorylation reaction would phosphorylate KaiC again.  The
free-energy landscape of the phosphorylation states combined with the
nucleotide binding pockets, where KaiA is not bound, is drawn in
\fref{CII_EnergyLandscape}, left panel.  We choose the magnitude of
the rates such that we can reproduce the time and height of the
maximum in the concentration of S-phosphorylated KaiC as well as its
subsequent decay.  Since dephosphorylation can only occur after the
ATP in the binding pocket has been hydrolyzed, the hydrolysis rate
constant sets an upper bound on the speed, which we set to
$\rmlabels{k}{CII}{hyd}=1$ h$^{-1}$, similar to what has been found in
\cite{Nishiwaki2012}.  \fref{CII_phospdyn}A shows that our model,
with the phosphotranfer parameters of \tref{CII_Parameters}, 
reproduces the dephosphorylation dynamics of Fig. S2 of \cite{Phong2012}. 
Next, we test if in our model the dephosphorylation
speed is independent of the bulk ATP fraction, $\alpha_{\rm ATP}={\rm
  [ATP]/([ATP]+[ADP])}$, as was found in experiments \cite{Rust2011}.
\eref{CII_phosphotransfer}C shows that this is indeed the case.  This
is because the ATP hydrolysis rate is higher than the ATP dissociation
rate and the phosphotransfer rate is faster than the ADP dissociation
rate.  This ensures that during dephosphorylation the nucleotides are
not released---if this would happen, the subsequent nucleotide binding
and hence the dephosphorylation rate
would depend on \rmlabels{\alpha}{}{ATP}.  When the above requirements
are fulfilled, KaiC predominantly dephosphorylates via
D$\cdot$ATP$\xrightarrow{\rmlabels{k}{CII}{hyd}}$D$\cdot$ADP$\xrightarrow{\rmlabels{k}{0}{DS}}$
S$\cdot$ATP$\xrightarrow{\rmlabels{k}{CII}{hyd}}$S$\cdot$ADP$\xrightarrow{\rmlabels{k}{0}{SU}}$U$\cdot$ATP.

\begin{figure}[ht!]
{\centering
\includegraphics[scale=1.0]{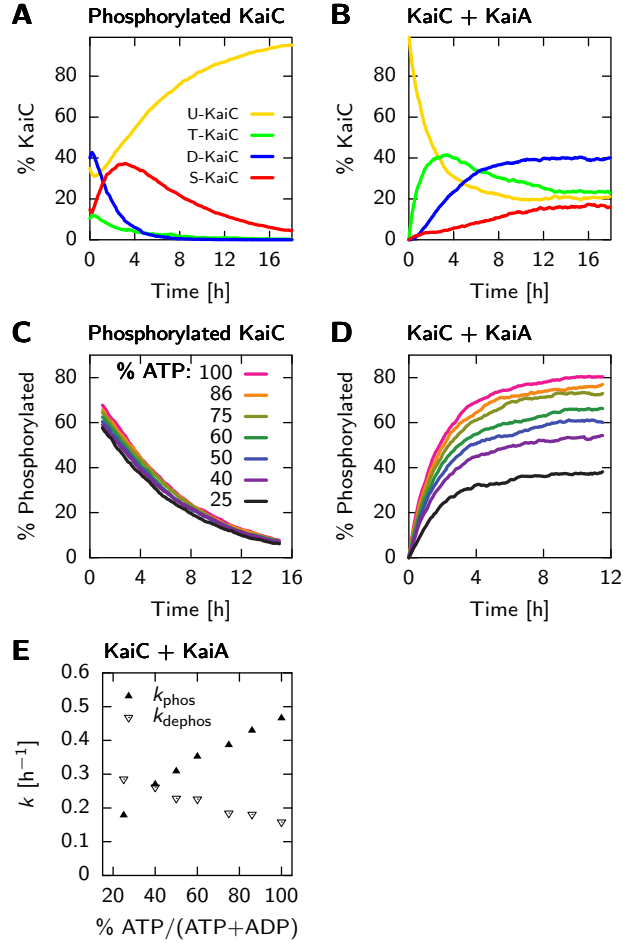}
\par
}
\caption{\flabel{CII_phospdyn} The free-energy landscape in \fref{CII_EnergyLandscape}, accurately describes the 
phosphorylation and dephosphorylation dynamics in the CII domain, both with and without KaiA and for a wide range 
of different ATP to ADP fractions in the solution, $\alpha_{\mathrm{ATP}}$.
(A) Dephosphorylation dynamics, starting with phosphorylated KaiC 
and (B) phosphorylation dynamics, starting with unphosphorylated KaiC and KaiA.
(C) Dephosphorylation and (D) phosphorylation of KaiC 
under similar conditions as in panels A and B, respectively, 
but now for different bulk fractions of ATP, $\alpha_{\mathrm{ATP}}$. 
Consistent with  experiments, the rate of dephosphorylation is independent of $\alpha_{\mathrm{ATP}}$, 
but the phosphorylation-rate does depend strongly on this fraction.
(E) Rates of (de)phosphorylation in panel D, 
found by fitting \eref{SimpleDephosModel} to the first four hours of phosphorylation data.
}
\end{figure}

{\bf Dephosphorylation does not occur via phosphate release.} 
Importantly, if dephosphorylation were to occur through the direct
release of inorganic phosphate groups into the bulk, 
then the dephosphorylation speed would \emph{also}
trivially be independent of the bulk ATP fraction.  We therefore set
out to test the hypothesis that all phosphorylation and
dephosphorylation occurs only through phosphotransfer  \cite{Nishiwaki2012, Egli2012}. To this end, we
compare
our model with an alternative one in which also direct exchange of the
phosphate group with the bulk is possible.  Specifically, we simulated
the experiments performed in \cite{Nishiwaki2012}, in which they track
radioactively labeled phosphate groups starting bound to the serine
and threonine sites of KaiC, in a solution with only non-radioactive
ATP.  Here, KaiC dephosphorylates while producing a
transient population of radioactive ATP$^{*}$, where at the maximum
around 20\% of the radioactive phosphates are bound to a nucleotide.
The inorganic phosphate groups, Pi$^{*}$, only appear in the bulk
after a marked delay.  \fref{CII_radPi}A shows that our model, where
dephosphorylation can only occur via phosphotransfer of the phosphate
to the ADP and the subsequent hydrolysis of the ATP, is in good
quantitative agreement with the results of Fig. 2 in \cite{Nishiwaki2012}:
Both the magnitude and timing of the peak in the radioactive ATP, ATP$^{*}$, 
and the delay in the appearance of radioactive phosphate groups
in the bulk, Pi$^{*}$, are in good agreement.

To test the effects of direct exchange of phosphate with the bulk, 
we add the reaction, ${\rm X\cdot N}\rates{}{}{\rm Y\cdot N + Pi}$, to our model.
Here X and Y are connected phosphorylation states and N is the state of the nucleotide binding pocket, 
which can be either ATP or ADP, but now does not change state during dephosphorylation.
We changed the rate constants such that dephosphorylation
occurs equally through both pathways and the overall speed is comparable to the original model.
\fref{CII_radPi}B shows that in this scenario, the time traces are qualitatively different from the experiment.
Because the phosphate groups are directly exchanged with the bulk,
the delay in [Pi$^{*}$] has disappeared and the magnitude of the peak in radioactive ATP is less than 10\%.
The discrepancy between the experimental data and our simulations
thus confirms our hypothesis that the direct exchange with the bulk of phosphates
is negligible.

\begin{figure}[ht!]
{\centering
\includegraphics[scale=1.0]{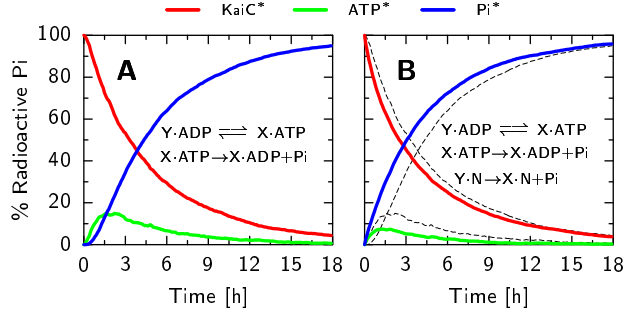}
\par
}
\caption{\flabel{CII_radPi} Tracking radioactive phosphate groups shows that 
  phosphotransfer between KaiC and ADP is the dominant pathway during dephosphorylation of KaiC. 
  Initial conditions are the same as in \fref{CII_phospdyn}A, but now the phosphate groups 
  are 'radioactive' such that they can be tracked from their initial position on the KaiC (KaiC$^{*}$, red line), 
  to the radioactive ATP (ATP$^{*}$, green line) and in solution (Pi$^{*}$, blue line). 
  Panels A show time traces of this dephosphorylation assay for our
  model, while panel B shows the results for an alternative model in which
  spontaneous dephosphorylation is also possible (for comparison, the
  results of our model, presented in panel A, are shown as thin dashed lines).
  In our model, panel A,
  dephosphorylation can only occur via the transfer of the phosphate
  group on KaiC to the ADP in the CII nucleotide binding pocket.  The
  alternative model, panel B, represents a scenario with two
  dephosphorylation pathways: one via the transfer to ADP, and one via
  direct exchange with the bulk.  We chose rates such that
  dephosphorylation occurs equally through both pathways, and the
  total dephosphorylation speed is similar to our model (panel A).
  Comparing with the radioactive phosphate tracking experiment in
  \cite{Nishiwaki2012}, Fig 2, shows that the scenario where
  dephosphorylation can only occur via ADP (panel A) agrees best with
  the data.  In particular, the onset of the free Pi concentration
  shows a temporal delay in panel A that is absent in panel B.
  Furthermore, the level of radioactive ATP is the highest in panel A,
  which compares best to the 20\% peak height in
  \cite{Nishiwaki2012}.}
\end{figure}

\subsubsection*{KaiA sets phosphorylation dynamics}
To address the effect of KaiA binding to the CII domain 
on the phosphorylation free-energy landscape shown in \eref{CII_DGKaiA},
we compared with the experimental time traces in Fig. S2 of \cite{Phong2012}.
In these experiments, they start with unphosphorylated KaiC together with KaiA and track the 
fractions of T,D and S phosphorylated KaiC.
This allows us to constrain the change in free energy of the phosphorylation states due to KaiA binding, 
$\delta\rmlabels{g}{CII\cdot KaiA}{bind}({\rm X})$, introduced in \eref{CII_DGKaiA}.
The affinity of KaiA for unphosphorylated KaiC is about 1nM, as reported in \cite{Qin2010}.
To make the large overshoot of T-phosphorylated KaiC possible, 
KaiA should lower the free-energy of the T state, $\delta\rmlabels{g}{CII\cdot KaiA}{bind}({\rm T})<0$,
while at the same time blocking the transition to the S-phosphorylation state, 
$\delta\rmlabels{g}{CII\cdot KaiA}{bind}({\rm S})>0$.
Experiments starting with radioactively labeled nucleotides in the binding pockets of KaiC in solution with KaiA,
show that ADP bound to the CII domain has a very high off-rate \cite{Nishiwaki-Ohkawa2014}.
Therefore, the ADP dissociation rate with KaiA bound to the CII domain has a high value of 
$\rmlabels{k}{CII\cdot ADP}{off, KaiA}=6.0$h$^{-1}$.
Taken together, we find that with the parameters for KaiA binding presented in \tref{CII_Parameters},
we can reproduce the phosphorylation dynamics as shown in \fref{CII_phospdyn}B.

{\bf Model can reproduce dependence phosphorylation on ATP fraction.}
Next we che\-cked the effect of the ATP fraction, $\alpha_{\rm ATP}$, 
on the speed and steady state level of the total phosphorylation fraction 
and compare with experiments in \cite{Rust2011}. 
As the sensitivity of phosphorylation to the bulk ATP fraction is set by 
the relative affinity for ATP and ADP, $\rmlabels{K}{CII}{ATP/ADP}$, 
we can use the data in \cite{Rust2011} to constrain this parameter.
Using $\rmlabels{K}{CII}{ATP/ADP}=0.10$, \fref{CII_phospdyn}D shows that the phosphorylation time traces 
at different values of $\alpha_{\rm ATP}$, are in good agreement with experiments:
Changing $\alpha_{\rm ATP}$ from 100\% to 25\%, 
the steady state phosphorylation level drops from 80\% to 40\%.

To quantify the change in phosphorylation speed by varying $\alpha_{\rm ATP}$, 
we fit the first 4 hours of the phosphorylation time traces in \fref{CII_phospdyn}D, $p(t)$,
with a 2 state model,
\begin{equation}
 p(t) = k_{\rm phos}/(k_{\rm phos} + k_{\rm dephos})\,(1-{\rm exp}(-(k_{\rm phos}+k_{\rm dephos})t).
 \elabel{SimpleDephosModel}
\end{equation}
Here, the system switches between the phosphorylated and dephosphorylated state with the
rates $k_{\rm phos}$ and $k_{\rm dephos}$.
\fref{CII_phospdyn}E shows that $k_{\rm phos}$ linearly increases with $\alpha_{\rm ATP}$,
as was found in \cite{Rust2011, Phong2012}, but with a slope that is less steep.
Furthermore, fitting the two state model to our modeling data yields a \rmlabels{k}{}{dephos} 
that decreases with \rmlabels{\alpha}{}{ATP}, 
while the fitting to the experimental data yields a \rmlabels{k}{}{dephos} that is virtually independent of 
\rmlabels{\alpha}{}{ATP}.
This decrease in our model is due to the fact that the effective dephosphorylation rate 
is proportional to the fraction of ADP in the CII binding pocket.
We attribute the inconsistency, at least in part, to the lower number of 
data points in the experimental time traces that are available for fitting to the two state model.

\subsubsection*{Ordered phosphorylation of the S and T sites persists in steady state}
The phosphotransfer reactions and the binding of KaiA all fulfill detailed balance,
and only the irreversible hydrolysis reaction in the CI and CII domains do not. 
This raises the question whether, when the solution contains only KaiC hexamers and KaiA, individual KaiC monomers
continue to go through the ordered cycle 
$\mathrm{U}\rightarrow\mathrm{T}\rightarrow\mathrm{D}\rightarrow\mathrm{S}\rightarrow\mathrm{U}$.
In this case there will be no macroscopic oscillations in the phosphorylation fraction 
because KaiA is never sequestered by KaiB and the phosphorylation cycles of the individual KaiC hexamers are not synchronized. 
The concentrations of the U,T,D and S phosphorylated monomers will therefore be in steady state.
If we find ordered phosphorylation of the threonine and serine sites, 
this has to be driven by the hydrolysis of ATP in the CI and/or the CII domain.
We want to know whether the phosphorylation cycle is mainly driven by hydrolysis in the CI or the CII domain.
To this end, below we consider scenarios where we remove the hydrolysis in the respective domains 
and study its effect on the phosphorylation dynamics.

To find out if the ordered phosphorylation cycle persist in a system with only KaiA and KaiC, 
we need to know if there are net fluxes between states in the phosphorylation state space, 
indicating that detailed balance is broken \cite{Kampen2007}.
To this end, we keep track of the number of phosphorylated threonine, 
\rmlabels{n}{}{T}(t), and serine residues, \rmlabels{n}{}{S}(t), 
in each individual hexamer in the ensemble.
From this data we can calculate the probability, $P_{\rmlabels{n}{}{T},\rmlabels{n}{}{S}}$, 
that a hexamer is in phosphorylation state $(\rmlabels{n}{}{T},\rmlabels{n}{}{S})$, 
and the number of times the hexamer switches
from this state to one of its neighboring states, $\rmlabels{N}{x}{\alpha, \beta}$.
Here $x\in\{T,S\}$ indicates whether the (de)phosphorylation event involved a threonine or serine residue,
$\alpha$ gives phosphorylation state $(\rmlabels{n}{}{T},\rmlabels{n}{}{S})$ before the transition, 
and $\beta$ the phosphorylation state after the transition.
% Here $x\in\{T,S\}$ indicates whether the (de)phosphorylation event involved a threonine or serine residue,
% $\alpha$ gives the number of phosphorylated sites of this type before the transition, 
% and $\beta$ the number after the transition.
The net flux between two neighboring states is given by
\begin{equation}
  \rmlabels{W}{x}{\alpha, \beta} = \frac{\rmlabels{N}{x}{\alpha, \beta}-\rmlabels{N}{x}{\beta, \alpha}}{\Delta \rmlabels{t}{}{sim}},
\end{equation}
where $\Delta\rmlabels{t}{}{sim}$ is the time interval over which these time traces were measured.
As was done in \cite{Battle2016}, we can define a vector, $\vec{J}_{\rmlabels{n}{}{T},\rmlabels{n}{}{S}}$, 
that points in the direction of the mean net flux through the state (\rmlabels{n}{}{T},\rmlabels{n}{}{S})
\begin{equation}
\vec{J}_{\rmlabels{n}{}{T},\rmlabels{n}{}{S}} = \frac{1}{2}
 \begin{pmatrix}
  \rmlabels{W}{T}{\alpha^{-}, \alpha} + \rmlabels{W}{T}{\alpha, \alpha^{+}} \\
  \rmlabels{W}{S}{\alpha^{-}, \alpha} + \rmlabels{W}{S}{\alpha, \alpha^{+}}
 \end{pmatrix}
 .
\end{equation}
Here the pair $\alpha^{-},\alpha$ indicates the net flux along the threonine or serine axis from below
the coordinate $\alpha$, and the pair $\alpha,\alpha^{+}$ indicates the net flux to above $\alpha$.

Hydrolysis of ATP in the CII domain provides the ADP required for dephosphorylation.
However, when there is no hydrolysis in the CII domain and the bulk only contains ATP, 
ATP is never converted to ADP in the CII domain (See \fref{CII_NucleotideExchange_ATPADP}),
such that dephosphorylation becomes impossible.  
All the monomers will be permanently in the D state with an ATP in the binding pocket,
blocking the possibility of a phosphorylation cycle. 
Therefore, in all scenarios discussed in this section, we will use a lower ATP fraction of $\alpha_{\rm ATP}=0.5$
such that there is ADP from the bulk available for dephosphorylation.

In \fref{CII_fluxes}, we show a heat map of $P_{\rmlabels{n}{}{T},\rmlabels{n}{}{S}}$ together
with the vectors $\vec{J}_{\rmlabels{n}{}{T},\rmlabels{n}{}{S}}$, for a system with KaiA and KaiC,
in the presence of ATP hydrolysis in both the CI and CII domains.
In panel A we show the behavior of KaiC in solution with KaiA,
which clearly shows cyclic net fluxes in phosphorylation state space. 
Starting in the lower left corner, where the hexamer is unphosphorylated, 
first the threonine sites will phosphorylate after which the serine sites become phosphorylated. 
After reaching the upper right corner of the state space, the hexamer will first dephosphorylate
the threonine sites and then the serine sites. 
As explained in the theory section on phosphorylation dynamics, 
the combination of hydrolysis in the CII domain and differential affinity for KaiA,
results in a high ATP fraction in the CII binding pockets when the hexamer is predominantly
in the U and T state, and a low ATP fraction when it is in the D and S state.
Therefore, KaiC, on average, phosphorylates when it is in the U and T state 
and dephosphorylates when it is in the D and S state, 
which is the origin of the cycle in $\rmlabels{n}{}{T}-\rmlabels{n}{}{S}$ space.

{\bf Cycle is driven by ATP hydrolysis and differential affinity.}
We then asked what drives the phosphorylation cycle at the level of the individual hexamers. 
We first set the hydrolysis rates in both domains to zero, 
thereby removing all irreversible pathways in the model. 
Clearly, without ATP hydrolysis, as shown in \fref{CII_fluxes}B, the fluxes disappear completely. 
This shows that our model obeys detailed balance when we remove the two irreversible pathways. 
When we remove hydrolysis in CI, $\rmlabels{k}{CI}{hyd}=0$, panel C, ordered phosphorylation persists, 
showing that hydrolysis in the CI domain is not essential for generating a cycle at the level of the individual hexamers.
If we only remove hydrolysis in CII and not in CI, 
there is still a small but clear cycle in state space due to the orchestrated switching between 
the active and inactive state caused by the phosphorylation state, 
and its effect on the affinity of the CII domain for KaiA (data not shown). 
Clearly, ATP hydrolysis in CI or CII is necessary to generate a cycle. 
Yet, it is not sufficient: If we remove differential affinity of KaiA binding to the CII domain 
but keep ATP hydrolysis in CI and CII, 
$\delta\rmlabels{g}{CII\cdot KaiA}{bind}(X_i)=0$ and $\Delta\rmlabels{G}{CII\cdot KaiA}{A,I}=0$, 
all fluxes disappear in phosphorylation state space, as shown in \fref{CII_fluxes}D. 
Hence, both differential affinity and ATP hydrolysis, most notably in the CII domain, 
are necessary to generate a phosphorylation cycle at the level of the individual hexamers. 

\begin{figure}[ht!]
\begin{center}
\includegraphics[scale=1.0]{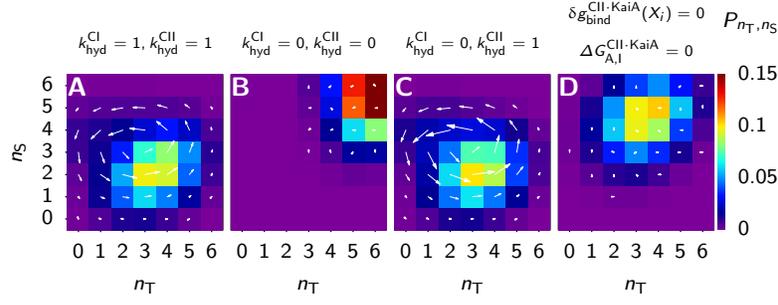}
\end{center}
\caption{\flabel{CII_fluxes} Hydrolysis in the CII domain, without KaiA sequestration by the CI domain, 
is sufficient to generate the ordered phosphorylation of the T and S sites in a solution of KaiC and KaiA.
In all panels, the bulk ATP fraction is 50\%.
(A-D) Heat maps of the probability, $P_{\rmlabels{n}{}{T},\rmlabels{n}{}{S}}$,  
for a single hexamer, of having $\rmlabels{n}{}{T}$ phosphorylated threonine sites
and $\rmlabels{n}{}{S}$ phosphorylated serine sites. 
Arrows indicate the net flux through a state, where the length is proportional to the magnitude of the flux.
(A) KaiC and KaiA, with hydrolysis of ATP in both the CI and CII domains, shows a clear ordered cycle in state space.
(B) Without hydrolysis in both the CI and CII domains, all the equations in our model obey detailed balance,
and the fluxes in state space disappear.
(C) When we remove hydrolysis in the CI domain, the fluxes in phosphorylation state space are little affected compared to panel A.
(D) When there is hydrolysis in both domains, but no differential affinity of the CII domain for KaiA,
the hydrolysis cycle in the CII domain does not couple to the phosphorylation cycle, and there is no ordered phosphorylation.
}
\end{figure}
 
\subsection*{Results on KaiC ATPase and cycle dynamics}
To test our model of the CI domain and to find the correct parameter values, 
shown in \tref{CI_Parameters}, we compare with the experiments on the
evolution of the ATP fraction in the binding pockets of KaiC and the
ATPase rate of KaiC \cite{Terauchi2007, Murakami2008, Nishiwaki2012, Nishiwaki-Ohkawa2014}. 
As is done in these experiments, we will
study the behavior of different combinations of the main actors: KaiA,
KaiB and KaiC and the ATP fraction \rmlabels{\alpha}{}{ATP}.  First we
will study a system containing only KaiC, which allows us to constrain
the parameters setting the rate of the hydrolysis cycle in the CI
domain.  Next, we look at the effect of dephosphorylation on the
transient ATP fractions in the CI and CII domains, which
provides an informative and testable prediction for how the CII domain regulates the
ATP fraction in CI, which is indeed one of the key characteristics of our model.  
Then we study the effect of KaiA on the steady
state ATPase rate and the dynamics of the ATP fraction in phosphorylating KaiC. 
For the full oscillating system, we will study
the phase difference between the phosphorylation level and ATPase
rate.  Then we study whether the oscillations are robust to changes in
the ATP fraction in solution, and whether our model can reproduce the
experimental observation that the period of the oscillation is
insensitive to these changes.  Again, all simulations were performed
with 720 KaiC hexamers and 720 KaiA dimers corresponding to the
experimental standard condition of a 0.6$\upmu$M concentration in a
volume of 2 cubic micron.

\begin{table}[ht!]
\begin{center}
    \begin{tabular}{| l | r l l | c |} \hline
     \multicolumn{5}{|c|}{}  \\[-7pt]
     \multicolumn{5}{|c|}{ \bf \large{Parameters relating to the CI domain}}  \\[3pt] \hline \hline
     \rowcolor{lightgray} Parameter & \multicolumn{3}{|c|}{\cellcolor{lightgray} Value } & Explanation \\ \hline 
     \multicolumn{5}{|c|}{\bf Nucleotide binding pocket} \\ \hline
     \rmlabels{k}{CI}{hyd} & 1.00 & h$^{-1}$ & & ATP hydrolysis in CI domain \\ 
     \rmlabels{k}{CI\cdot ADP}{off} & 1.50 & h$^{-1}$ & &  ADP off-rate in CI domain \\ \hhline{|~|~~~|-|}
     $\delta \rmlabels{g}{CI\cdot ADP}{act, A}(U)$ &  0.00 &kT & & Activation energy contributions \\ 
     $\delta \rmlabels{g}{CI\cdot ADP}{act, A}(T)$ & -0.80 &kT & & from the respective monomers, \\ 
     $\delta \rmlabels{g}{CI\cdot ADP}{act, A}(D)$ &  0.40 &kT & & in the active state. \\ 
     $\delta \rmlabels{g}{CI\cdot ADP}{act, A}(S)$ &  0.80 &kT & &  \\ \hhline{|~|~~~|-|}
     $\delta \rmlabels{g}{CI\cdot ADP}{act, I}(U)$ & -0.20 &kT & & Activation energy contributions \\ 
     $\delta \rmlabels{g}{CI\cdot ADP}{act, I}(T)$ & -0.80 &kT & & from the respective monomer, \\ 
     $\delta \rmlabels{g}{CI\cdot ADP}{act, I}(D)$ &  0.40 &kT & & in the inactive state. \\ 
     $\delta \rmlabels{g}{CI\cdot ADP}{act, I}(S)$ &  0.80 &kT & &  \\ \hline \hline    
     \multicolumn{5}{|c|}{\bf KaiA and KaiB sequestration dynamics } \\ \hline
     \rmlabels{k}{CI\cdot KaiA}{on, A}          & 1.00 & $\cdot 10^{6}$ & $\upmu$Mh$^{-1}$ & KaiA on-rate, active \\ 
     \rmlabels{k}{CI\cdot KaiA}{off, A}         & 1.00 & $\cdot 10^{1}$ & h$^{-1}$ & KaiA off-rate, active \\ 
     \rmlabels{k}{CI\cdot KaiA}{on, I}  & 1.00 & $\cdot 10^{6}$ & $\upmu$Mh$^{-1}$ & KaiA on-rate, inactive \\ 
     \rmlabels{k}{CI\cdot KaiA}{off, I} & 1.00 & $\cdot 10^{-1}$ & h$^{-1}$ & KaiA off-rate, inactive \\ 
     \rmlabels{k}{CI\cdot KaiB}{on, A}          & 1.00 & $\cdot 10^{-1}$ & h$^{-1}$ & KaiB on-rate, active \\ 
     \rmlabels{k}{CI\cdot KaiB}{off, A}         & 1.00 & $\cdot 10^{1}$ & h$^{-1}$ & KaiB off-rate, active \\ 
     \rmlabels{k}{CI\cdot KaiB}{on, I}  & 2.00 & $\cdot 10^{0}$ & h$^{-1}$ & KaiB on-rate, inactive  \\
     \rmlabels{k}{CI\cdot KaiB}{off, I} & 1.00 & $\cdot 10^{-2}$ & h$^{-1}$ & KaiB off-rate, inactive \\ 
     \rmlabels{n}{CI\cdot KaiA}{max}            & 6                  & &          & \#KaiA sequestered/hexamer \\ 
     \rmlabels{n}{CI\cdot KaiB}{max}            & 6                  & &          & \#KaiB sequestered/hexamer \\ \hline \hline
     \multicolumn{5}{|c|}{\bf Conformational state } \\ \hline
     \rmlabels{k}{conf}{0}                   & 10 		    & & h$^{-1}$ & prefactor conformational switch \\ 
     \rmlabels{n}{CI\cdot ADP}{0}            &  5 		    & &          & Offset energy in conformation \\ 
     $\delta \rmlabels{g}{ATP,ADP}{A,I}$       & 19 		    & & kT     & A,I ADP dependent energy \\ \hline
    \end{tabular}    
\end{center}
\caption{\tlabel{CI_Parameters} Model parameters relating to the CI domain
are introduced in the theory section on the power cycle and their values are motivated in the results section.
Energies are given in units of kT, were k is Boltzmann's constant and T the temperature.
Note that the rates of binding and unbinding of KaiA are for the CI domain. 
For rates relating to the CII domain, see \tref{CII_Parameters}.
}
\end{table}

\subsubsection*{Ensemble of dephosphorylating KaiC shows a transient decrease in ATPase rate}
To constrain the hydrolysis rate constant and the ADP dissociation rate in the CI domain of unphosphorylated KaiC, 
we used the experimental observation that in a system with only KaiC that has reached steady state,
the ATP fraction in the binding pockets is around 30\% \cite{Nishiwaki-Ohkawa2014}
and the ATPase rate is 0.6 ATP per KaiC monomer per hour \cite{Terauchi2007}.
As we argued in the theory section on KaiA acting as a nucleotide exchange factor, 
without KaiA, the CII binding pockets are predominantly occupied by ADP,
the monomers are in the U state,
and the ATPase activity comes mainly from hydrolysis in the CI domain.
Now, since the total fraction of ATP in the binding pockets is given by 
0.5(\rmlabels{\beta}{CI}{ATP}+\rmlabels{\beta}{CII}{ATP})=0.3, 
where \rmlabels{\beta}{CI}{ATP} and \rmlabels{\beta}{CII}{ATP} are the ATP fractions
in the CI and CII domain, respectively,
we estimate that the fraction of ATP in the CI domain, \rmlabels{\beta}{CI}{ATP}=0.6,
because \rmlabels{\beta}{CII}{ATP}$\approx$0.
Assuming the measured ATPase rate equals the hydrolysis rate constant 
times the fraction of ATP in the CI binding pocket, 
\rmlabels{k}{CI}{hyd}\,\rmlabels{\beta}{CI}{ATP}, we estimate that \rmlabels{k}{CI}{hyd}=1.0 h$^{-1}$.
Since \rmlabels{\beta}{CI}{ATP} = \rmlabels{k}{CI\cdot ADP}{off}/(\rmlabels{k}{CI\cdot ADP}{off} + \rmlabels{k}{CI}{hyd}), 
we deduce that \rmlabels{k}{CI\cdot ADP}{off}=1.5 h$^{-1}$.

To find out if an ensemble of only KaiC hexamers has the observed dynamics of the ATP fraction,
we first consider a system in which KaiC is unphosphorylated and the ATP fraction in the binding pockets 
is 100\%.
We then study its relaxation to steady state.
\fref{CICII_ATPase}A shows an exponential decay of the ATP fraction in both domains,
on a similar timescale and steady state value as was found in \cite{Nishiwaki2012}.
The mean ATP fraction and ATPase rate, given in \tref{CICII_ATPase}, 
are in quantitative agreement with experimental data presented above.

\begin{table}[ht!]
\begin{center}
    \begin{tabular}{ c c l | r  r  r  r } 
     \multicolumn{3}{c}{Mixture} & \multicolumn{4}{c}{$\langle${ATPase}$\rangle$  [\#ADP/KaiC/day]} \\ \hline \hline
     KaiA & KaiB & \multicolumn{1}{c|}{Condition} & \multicolumn{1}{c}{CI} & \multicolumn{1}{c}{CII} & \multicolumn{1}{c}{CI+CII} & From \cite{Terauchi2007} \\ \hline
     - & - &                                  & 14.4 & 0.0  & 14.4 & 14.5 \\ 
     + & - &                                  & 11.5 & 17.0 & 28.5 & 18.1 \\ 
     - & + &                                  & 14.2 & 0.0  & 14.2 &  8.9 \\ 
     + & + &                                  & 12.3 &  8.8 & 21.1 & 15.8 \\
     + & + & $\alpha_{\mathrm{ATP}}=50\%$     & 13.0 &  8.3 & 21.3 & -    \\
     + & - & \rmlabels{k}{CII\cdot KaiA}{hyd} = 0 & 10.8 & 4.2 & 15.0 & -  \\
     + & + & \rmlabels{k}{CII\cdot KaiA}{hyd} = 0 & 13.9 &  3.0 & 16.9 & - \\ 
     + & - & \rmlabels{K}{CII}{ATP/ADP}(D)=10 & 11.2 & 11.4 & 22.6 & -    \\
     + & + & \rmlabels{K}{CII}{ATP/ADP}(D)=10 & 12.8 &  7.4 & 20.2 & -   \\ \hline
    \end{tabular}
\end{center}
\caption{\tlabel{CICII_ATPase} Measured ATPase activity in ADP molecules produced per 
KaiC monomer per day (24 hours), under different conditions.
First rows show results for parameters given in \trefs{CII_Parameters} and \ref{table:CI_Parameters}.
The last four rows show results for alternative models.
When $\rmlabels{k}{CII\cdot KaiA}{hyd}=0$, the ATP hydrolysis in the CII domain is blocked when KaiA is bound to CII.
When, $\rmlabels{K}{CII}{ATP/ADP}=10$, the CII domain has a higher relative affinity for ADP 
when the monomer is in the D state.
The experimental values for the combined ATPase activity of the CI and CII domain, given in the last column, are taken from \cite{Terauchi2007}, 
and are shown for comparison.
}
\end{table}

Next we want to find out how the ATP fraction evolves in a system in which KaiC is initially 
highly phosphorylated, with the concentrations of monomers 
in the U,T,D and S state evolving as shown in \fref{CII_phospdyn}A.
Please note that in our model, 
the T state will lower the activation energy for ADP dissociation set by 
$\delta\rmlabels{g}{CI\cdot ADP}{barrier, A/I}(X_i)$,
while the D and S state will increase the activation energy.

The parameter values for the contributions to the activation energy,
$\delta\rmlabels{g}{CI\cdot ADP}{barrier, A/I}(X_i)$, given
in \tref{CI_Parameters}, were chosen to give good agreement of the
oscillation dynamics presented below.  We predict that the peak in the
number of monomers in the S state\linebreak (\fref{CII_phospdyn}A), which will
decrease the ADP dissociation rate, causes a transient lowering of the
overall bound ATP fraction, particularly in the CI domain.  Indeed,
\fref{CICII_ATPase}B shows a clear trough in the fraction of ATP in
the binding pockets, which is most pronounced for CI.  An
experiment tracking the ATP fraction in the binding pockets, as
performed in \cite{Nishiwaki-Ohkawa2014}, but now starting with
phosphorylated monomers, would be able to verify this
prediction. Lastly, in our model the ATPase rate is proportional to
the fraction of ATP in the binding pockets. Our model thus predicts a
transient dip in the ATPase rate for dephosphorylating KaiC hexamers.

\begin{figure}[ht!]
{\centering
\includegraphics[scale=0.95]{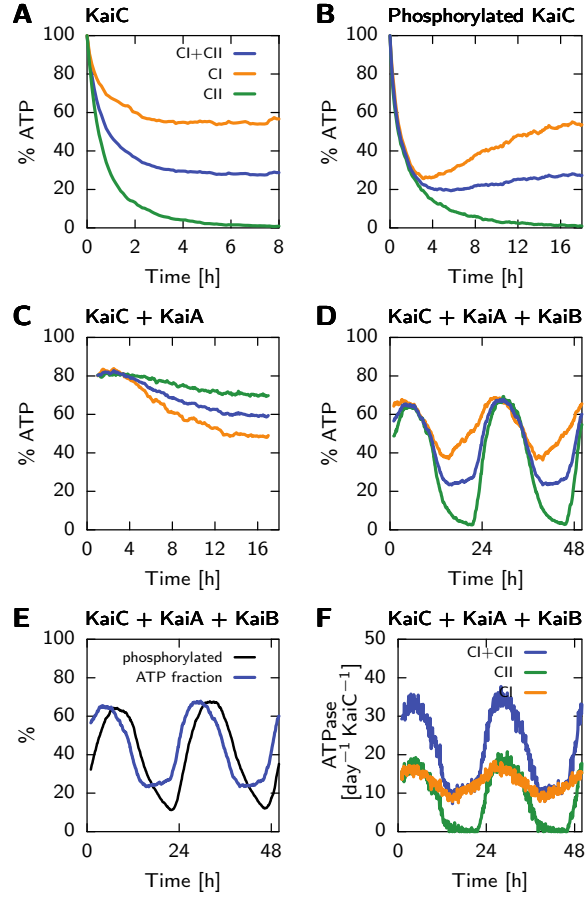}
\par
}
\caption{\flabel{CICII_ATPase} Model correctly predicts the ATP
  fractions in KaiC nucleotide binding pockets and the phase
  difference between this fraction and the phosphorylation level.
  Figures A-D show the fraction ATP/(ATP+ADP) in the nucleotide binding pockets of the CI
  domain (orange), the CII domain (green) and their sum (blue), in a
  100\% ATP solution, for different scenarios: KaiC initially
  unphosphorylated, no KaiA and KaiB present (A), KaiC initially
  phosphorylated, no KaiA and KaiB (B), initially unphosphorylated
  KaiC + KaiA (C) and KaiC + KaiA + KaiB (D).  (A) The ATP levels drop
  monotonically due to the slow hydrolysis in both the CI and CII domain. 
  (B) ATP fractions in dephosphorylating KaiC shows a clear trough in the ATP fraction of
  the CI domain due to the peak in the number of monomers in the S
  state (\fref{CII_phospdyn}A), which temporarily decreases the
  ADP-off-rate in the CI domain.  (C) For a system with KaiC and KaiA,
  the ATP fraction is higher in the CII domain and lower in the CI
  domain.  KaiA increases the nucleotide exchange rate in the CII
  domain, and the resultant high phosphorylation level decreases the
  ADP dissociation rate in CI, which decreases the ATP fraction in
  this domain.  (D) Full system with KaiA, KaiB and KaiC shows
  oscillations in the ATP fractions of both the CI and CII domains.
  (E) The phase difference and amplitude of the ATP fraction 
  (\% ATP, blue line) 
  and the phosphorylation level (\% phosphorylated monomers, black line) agree well with
  experimental results in \cite{Nishiwaki-Ohkawa2014}.  
  (F) ATPase levels of the KaiC domains show oscillations proportional 
  to the ATP fractions in the respective domains.  }
\end{figure}

\subsubsection*{KaiA has opposite effects on the ATP fraction in the CI and CII domains}
Adding KaiA to an ensemble of KaiC will immediately increase the ATP
fraction in the CII binding pockets, causing phosphorylation of KaiC
as shown in \fref{CII_phospdyn}B.  Due to the phosphorylation of KaiC,
the ATP fraction in the CI domain will decrease, because the ADP
release in CI depends on the phosphorylation state of CII.  These
results are illustrated in \fref{CICII_ATPase}C, which also shows that
the total ATP fraction in CI and CII stabilizes around 60\%, in good
agreement with \cite{Nishiwaki-Ohkawa2014}.  However, the steady state
ATPase rate of the ensemble is around 29 ADP molecules produced per
monomer per day (ADP/KaiC/day), \tref{CICII_ATPase}, higher than the
observed rate of 18 ADP/KaiC/day \cite{Terauchi2007}. This descrepancy
could be due to our assumption that the ATP hydrolysis rate is
constant, independent of both the phosphorylation state and whether or
not KaiA is bound, and/or the assumption that the nucleotide
affinities in the CII domain are independent of the phosphorylation
state.  It is conceivable that KaiA, when bound to CII, decreases the
ATP hydrolysis rate in the CII domain. When we set the hydrolysis rate
to zero when KaiA is bound, \rmlabels{k}{CII\cdot KaiA}{hyd}=0, and
adjust the ADP dissociation rate to keep the phosphorylation dynamics
unchanged, \rmlabels{k}{CII\cdot ADP}{off, KaiA}=0.2, we find that the
steady state ATPase rate in the system with only KaiA and KaiC drops
to 15 ADP/KaiC/day (see \tref{CICII_ATPase}), in good agreement with
experiment. Another possibility is indeed that the affinity of doubly
phosphorylated KaiC for ATP is lower than assumed in our model. In
our current model, doubly phosphorylated KaiC has the same high affinity for
ATP as KaiC in the other phosphorylation states (see \fref{CII_EnergyLandscape}). 
When only KaiA is present and KaiC is often in the doubly phosphorylated D state, 
this sets up a futile cycle, in which KaiC continually binds ATP and then hydrolyzes it. 
Decreasing the relative affinity of D for ATP versus ADP, such that $\rmlabels{K}{CII}{ATP/ADP}=10$, 
lowers the steady state ATPase rate to about 22 ADP/KaiC/day. 
By making minor adjustments to parameters of the model, 
our model can thus reproduce the ATPase rate of KaiC in the presence of KaiA. 

\subsubsection*{Delayed KaiA sequestration synchronizes the hexamers}
Given our analysis in the results section on the phosphorylation dynamics, 
which showed that KaiC goes through a ordered phosphorylation cycle in solution with KaiA, 
we wanted to know if the sequestration of KaiA, 
after the slow binding of KaiB, will synchronize the KaiC hexamers. 
To this end we chose rates for KaiB binding and unbinding as presented in
\tref{CI_Parameters}, which correspond to a very low affinity for KaiC
in the active state, where KaiB is almost never bound, and a high
affinity for the inactive state where KaiB binds and unbinds slowly.
KaiA binds rapidly to the inactive CI domain when 6 KaiB monomers are bound to it, 
and dissociates from this complex very slowly.
The last important quantity relating to the CI domain, 
which determines the stabilization of the inactive state of KaiC by ADP in the CI domain,
$\delta\rmlabels{g}{ATP,ADP}{A,I}$, is constrained from below by the
relative affinities of KaiB and KaiA for the inactive state compared
to the active state, as discussed in the theory section on the power cycle.
Given the parameters for KaiA and KaiB binding to the CI domain in
\tref{CI_Parameters}, we choose $\delta\rmlabels{g}{ATP,ADP}{A,I}=19$
kT.

\frefs{CICII_ATPase} D,E and F show clear oscillations in the ATP fraction in the CI and CII binding pockets,
the phosphorylation fraction and the ATPase rates of the Kai oscillator, respectively.
Panel E shows that the phase of the phosphorylation fraction is a few hours ahead of the 
ATP fraction in the binding pockets, which is in good agreement with experiments \cite{Nishiwaki-Ohkawa2014}.
Also the amplitudes of both oscillations are in good agreement.
The average ATPase rate of the oscillator is about 21 ADP/KaiC/day, 
which is slightly higher than the observed rate of 15 ADP/KaiC/day. 
We hypothesize that this high ATPase activity has to be attributed to the ATP hydrolysis in the CII domain.
To test this, we set, as in the previous section, the hydrolysis rate
constant to zero, $\rmlabels{k}{CII\cdot KaiA}{hyd}=0.0$,
when KaiA is bound to the CII domain.
In order for the phosphorylation dynamics to be comparable to our original model,
we set $\rmlabels{k}{CII\cdot ADP}{off, KaiA}=0.2$.
In this model, the average ATPase activity has dropped to 16 ADP/KaiC/day,
almost equal to the experimentally observed value (\tref{CICII_ATPase}). 
Decreasing the affinity of doubly phosphorylated KaiC for ATP,
which strongly reduced the ATPase rate of KaiC in the presence of
KaiA only (see previous section), has a much smaller effect when both KaiA and
KaiB are present, lowering the ATPase rate to 20 ADP/KaiC/day.

\subsubsection*{Clock period independent of bulk ATP fraction}
To find out how robust our model of the Kai oscillator is against
changes in the steady-state ATP level, we simulate the oscillator at
different ATP fractions, \rmlabels{\alpha}{}{ATP}, as was done
experimentally in \cite{Phong2012}.  \frefs{CICII_cT} A,B and C show
that the time traces of concentrations of monomers in the T,D and S
state with $\rmlabels{\alpha}{}{ATP}=1.0,0.75$ and 0.50, respectively,
are in good quantitative agreement with experiments.  Both the
amplitude of the concentrations and their relative phases agree.
However, contrary to experiments, at $\rmlabels{\alpha}{}{ATP}=0.25$
the ensemble does not oscillate anymore.  The reason is that not
enough hexamers bind 6 KaiB monomers to sequester all the KaiA.  Since
the absence of KaiA normally synchronizes the phosphorylation state of
all hexamers, now the ensemble quickly becomes unsynchronized, and the
oscillations disappear. Interestingly, in the range of ATP fractions
where the Kai system does oscillate, the period of the oscillations is
independent of \rmlabels{\alpha}{}{ATP}, as is shown in panel E, and
experimentally observed in \cite{Rust2011}. 
This is remarkable, because we did not design the system to have a period
independent of \ATPfrac. In panel F we plot the peak, 
trough and mean phosphorylation level of the oscillations at
different values of \rmlabels{\alpha}{}{ATP}.  The increase in peak
height with \rmlabels{\alpha}{}{ATP} is in good agreement with
experiments in \cite{Phong2012}.  However, while in our simulations
the level of the troughs only marginally rises as
\rmlabels{\alpha}{}{ATP} decreases, experiments show a considerable
increase.  Because in our simulations the appearance of free KaiA at
the end of the cycle always occurs at a fixed phosphorylation ratio,
it is hard to explain this discrepancy with experiments.

\begin{figure}[ht!]
{\centering
\includegraphics[scale=1.0]{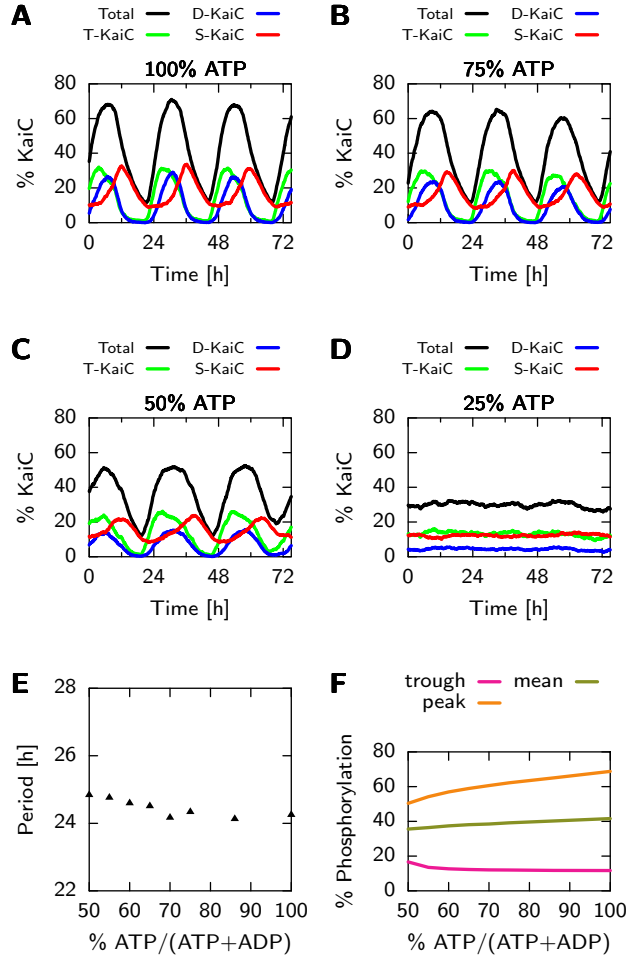}
}
\caption{\flabel{CICII_cT} The Kai system oscillates over a wide range of ATP fractions, 
while the period remains unchanged.
(A-D) Time traces of monomers in the T state (green), D state (blue), 
S state (red) and total phosphorylation level (black), for different ATP fractions in the bulk. 
Amplitudes and phases in good agreement with experiment (Compare with \cite{Phong2012}, Fig. S1), 
except that our system does not oscillate at 25\% ATP fractions.
(E) The average peak-to-peak time of the phosphorylation fraction for a 1000 hour time trace at different
ATP fractions. The period is remarkably unaffected by the ATP fraction, even though it has a big influence 
on the phosphorylation speed, as we showed in \fref{CII_phospdyn}, D. 
(F) Mean peak and trough phosphorylation levels at different ATP fractions. 
The increase in peak-hight with \rmlabels{\alpha}{}{ATP} is in in good agreement with experiment. 
However, in our simulations, the level of the trough is almost independent of \rmlabels{\alpha}{}{ATP}, 
while in experiments it increases with \rmlabels{\alpha}{}{ATP} (Compare with Fig. S6 in \cite{Rust2011} and Fig. 1 in \cite{Phong2012})
}
\end{figure}

\subsubsection*{What is the principal driver of the oscillations: hydrolysis in the CI or the CII domain?}
In our model, there are two reactions that break detailed balance and allow the system to oscillate: 
Hydrolysis of ATP in the CI and in the CII domain.  
In the results section on phosphorylation dynamics we showed that hydrolysis of ATP in CII, 
in combination with differential affinity, is sufficient to generate cycles of phosphorylation 
at the level of the individual hexamers. 
However, without hydrolysis of ATP in the CI domain, the macroscopic oscillations inevitably come to a halt, 
because not enough KaiC can reach the inactive conformational state to allow for the necessary level of periodic KaiA sequestration.
Clearly, while ATP hydrolysis in CI is not essential for generating cycles at the level of individual hexamers, 
it is necessary for generating macroscopic oscillations. 
The question that remains is whether hydrolysis of ATP in CII is likewise necessary for creating coherent, 
macroscopic oscillations. 
In the model presented so far, hydrolysis of ATP in CII is needed to allow for dephosphorylation: 
During the dephosphorylation phase, when a KaiC protein has made the transition from the D to the S state, 
it has ATP in the binding pocket, which needs to be hydrolyzed to generate ADP, 
thereby enabling the transition from S to U. 
However, this ATP hydrolysis reaction does not seem of fundamental importance 
for breaking detailed balance and creating macroscopic oscillations. 
Could this system generate oscillations with only turnover of ATP in the CI domain? 
To address this question, we here investigate a slightly modified version of our original model 
that has no ATP hydrolysis in the CII domain.
This modified version does not represent the real Kai system, 
but rather serves as a thought experiment to clarify the different thermodynamic roles 
of ATP hydrolysis in the two domains.

We change our existing model so that KaiC can dephosphorylate without hydrolysis in the CII domain
while still having KaiA stimulated phosphorylation in order to synchronize the hexamers.
To this end, we set the ATPase rate in the CII domain to zero, 
while still allowing phosphates to be transferred in both directions between ATP (or ADP) 
and the serine and threonine residues.  
For dephosphorylation to occur, 
there must then be some mechanism other than ATP hydrolysis to introduce ADP's 
into the CII binding pocket to receive the phosphates.  
We thus make the nucleotide exchange rate high in the inactive state.
In the active state, however, 
the nucleotide exchange rate should still be low unless KaiA is bound to CII,
to preserve the mechanism of KaiA stimulated phosphorylation.  
To maintain the correct relative stability of the two KaiC conformations, 
ADP in the CII domain now also stabilizes the inactive state,
such that ADP has a high affinity for the CII domain when the hexamer is in the inactive state,
\rmlabels{\tilde{K}}{CII}{ATP/ADP}=10.0,
and the original low relative affinity for ADP when in the active state \rmlabels{K}{CII}{ATP/ADP}=0.10.
Finally, because ADP from the bulk is required for dephosphorylation, we set \ATPfrac=0.5.
Changed parameters are listed in \tref{NewModelParameters} below.
\begin{table}[h!t]
 \begin{tabular}{l r | r r}
  Parameter &  & Active & Inactive \\ \hline \hline
  \rmlabels{k}{CII}{hyd}                & (h$^{-1}$) & 0 & 0  \\
  \rmlabels{K}{CII}{ATP/ADP}            &            & 0.1 & 10 \\
  \rmlabels{k}{CII\cdot ATP}{off, 0}    & (h$^{-1}$) & 0.6 & 6.0 \\
  \rmlabels{k}{CII\cdot ATP}{off, KaiA} & (h$^{-1}$) & 6.0 & 6.0 \\ 
  \rmlabels{g}{ATP/ADP}{A,I}            & (kT)       & 30 & 30 \\
 \end{tabular}
 \caption{\tlabel{NewModelParameters} Parameters used in the alternative model without hydrolysis in the CII domain
 that are different from the values in our original model in Tables \ref{table:CII_Parameters} and \ref{table:CI_Parameters}. 
Parameters values are listed for both the active and inactive conformations 
if they have been changed for either conformation.}
\end{table}

Our modified model shows robust macroscopic oscillations, as shown in \fref{PrincipleDriver}A.
The total ATP consumption has dropped to 11.6 ATP/KaiC/day, all due to the CI domain's ATPase activity.
This shows that the Kai oscillator can in principal generate macroscopic oscillations with only
hydrolysis in the CI domain, and that hydrolysis in CII is not essential.
However, as shown in \fref{PrincipleDriver}B, 
the rate of dephosphorylation now strongly depends on the ATP fraction in the bulk, 
because \ATPfrac{} affects the probability that, upon the D$\rightarrow$S transition and subsequent ATP release, 
CII will bind ADP instead of ATP, which is necessary for the next S$\rightarrow$U transition. 
This dependence of the dephosphorylation rate on the bulk ATP fraction
in this modified model is contrary to what is observed in experiments.
Even when we allow for hydrolysis in the CII domain
in addition to nucleotide exchange in the inactive state, \fref{PrincipleDriver}C,
the dephosphorylation speed still strongly depends on \ATPfrac.
As argued in \cite{Phong2012}, 
such a dependence of the dephosphorylation rate on \ATPfrac{} would hamper input compensation,
and the period of the oscillations would not be constant any more at different ATP fractions.
A low nucleotide exchange rate in the inactive state 
(as included in our main model described in the preceding sections) 
seems therefore critically important for the real Kai oscillator.

\begin{figure}[ht!]
\begin{center}
\includegraphics[scale=1.0]{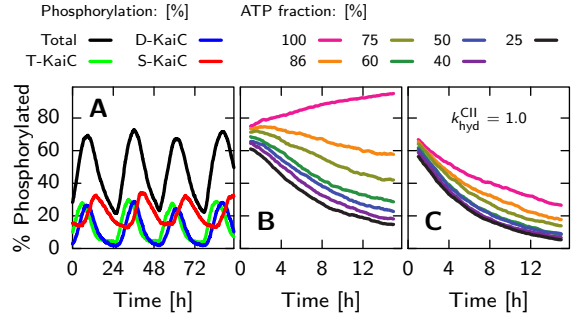} 
\end{center}
\caption{\flabel{PrincipleDriver} An alternative model, without hydrolysis in the CII domain,
can generate robust macroscopic oscillations.
(A) Time traces of monomers in the T state (green), D state (blue), S state (red) 
and total phosphorylation level (black), at 50\% ATP level in the bulk,
for the alternative model with hydrolysis only in the CI domain.
(B) Dephosphorylation of KaiC for different bulk fractions of ATP, \ATPfrac{}. 
Due to the high nucleotide exchange rate of the CII domain in the inactive conformation,
the dephosphorylation speed becomes sensitive to \ATPfrac{},
contrary to what is observed experimentally.
(C) Even when we add the hydrolysis of ATP in the CII domain, $\rmlabels{k}{CII}{hyd}=1$,
the speed still depends on \ATPfrac{}.}
\end{figure}

\section*{Discussion}
We set out to develop a thermodynamically correct model of the
post\hyp{}translational Kai oscillator that is consistent with the
large body of quantitative experimental data available. 
In particular, the recent experimental observation that KaiC regenerates
ATP during dephosphorylation \cite{Nishiwaki2012, Egli2012} made us
rethink the thermodynamics behind the phosphorylation cycle: If
phosphorylation and dephosphorylation require no net turnover of ATP,
what drives the thermodynamic cycle of the oscillator? 
We built our new model on our earlier model of the Kai system \cite{VanZon2007, Zwicker2010}, 
where each individual KaiC hexamer goes through a
cycle of phosphorylation and dephosphorylation. 
The hexamers are synchronized through the mechanism of differential affinity, 
in which the affinity of KaiA for KaiC changes with the phosphorylation state
and the complex of inactive KaiC with KaiB has the strongest affinity for KaiA; 
sequestration of KaiA by this complex allows the hexamers lagging behind and which are 
still in the dephosphorylation phase of the cycle to remove KaiA from the front runners, 
which have already completed their cycle yet need KaiA to be phosphorylated again. 
Here, we have extended these ideas with a more detailed monomer model, 
taking into account the two phosphorylation sites per
monomer and the nucleotide binding pockets in the CI and CII domain.
This allowed us to include the effects of the ordered phosphorylation
cycle on the monomer level, found in \cite{Rust2007, Phong2012}.

\subsection*{Summary of the key results of the model}
Here we give an overview of the most important conclusions we can draw from our model.

{\bf By enhancing nucleotide exchange, KaiA both stimulates phosphorylation 
and blocks dephosphorylation, preventing futile cycles.} 
In our new model, next to the threonine and serine
phosphorylation sites, we explicitly track whether there is ATP or ADP
present in the nucleotide binding pockets of the CI and CII domains of
KaiC.  Dephosphorylation proceeds exclusively by phosphotransfer
between the phosphorylation site and ADP in the CII domain; direct
release of inorganic phosphate from serine or threonine residues to the bulk can not occur. 
KaiA acts as a nucleotide exchange factor \cite{Nishiwaki-Ohkawa2014}, 
enhancing the release rates of nucleotides
from the CII domain.  In combination with the hydrolysis of ATP in the
CII domain, KaiA effectively increases the fraction of ATP in the
binding pocket, thereby driving the hexamer towards the highly
phosphorylated state.  Because dephosphorylation with ATP in the
binding pocket is impossible, KaiA, when bound, also effectively blocks
dephosphorylation.  In this way, the serine and threonine sites can be
phosphorylated with a rate similar to the rate of dephosphorylation,
because the phosphorylation rate does not need to compensate for the
spontaneous release of phosphate groups.  The capacity of KaiA to
block dephosphorylation has two important consequences. First, if KaiA
were not to block dephosphorylation, then the latter would inevitably
occur, which must then be followed by another round of
phosphorylation. Such futile cycles would make the oscillator less
efficient.  Furthermore, if KaiA were not to impede dephosphorylation,
then {\em net} phosphorylation could only occur if the phosphorylation rate is
larger than the dephosphorylation rate. Simulations reveal, however,
that in such a model the concentration of threonine phosphorylated
KaiC would rise too fast compared to the phosphorylation assay shown
in \fref{CII_phospdyn}B. The timing of the peak of T state monomers
would then be wrong.

%{\bf KaiA is essential for ordered phosphorylation of the S and T sites.}
{\bf Differential affinity of the CII domain for KaiA stimulates 
the ordered phosphorylation of the S and T sites.}
Next to regulating the ATP level in CII binding pockets, KaiA
also steers the order of phosphorylation of the threonine and serine
sites, which arises in our model as a logical consequence of
differential affinity.  In our model, KaiA has a higher affinity for
KaiC that is in the T state, with threonine phosphorylated, than in
the S state, with serine phosphorylated. Detailed balance then implies
that the binding of KaiA raises the energy level of the S state
compared to that of the T state (see the energy levels of
\fref{CII_EnergyLandscape}).  In this way, the binding of KaiA to CII
drives the ordered phosphorylation cycle, where first the threonine
site is phosphorylated and then the serine site.  We tested this with
simulations, and found that KaiC hexamers in a solution with KaiA and
no KaiB, go through the ordered phosphorylation cycle of the T and S
sites.  Hydrolysis of ATP bound to the CII domain is sufficient to
drive this cycle. This prediction could conceivably be tested by
  performing mass-spectrometry experiments \cite{Rust2007, Murayama2011},
  tracking the phosphorylation states starting from different initial
  conditions with synchronized hexamers, and testing whether the
  measured rates of phosphorylation and dephosphorylation obey
  detailed balance.

%Even though kaiB is not essential for the ordered phosphorylation cycle in our model,
  {\bf ATP hydrolysis in CI drives conformational switching and macroscopic oscillations.} 
  Even though hydrolysis in the CII domain is sufficient to give rise to the
  phosphorylation cycle of the {\em individual} hexamers, we
  conclude that this is not the principal driver of the {\em
    macroscopic} oscillations, and in particular of the periodic
    sequestration of KaiA.  If the phosphorylation of KaiC were to
  directly stabilize the inactive state and consequently the binding
  of KaiB, then detailed balance would dictate that, conversely, the
  latter also stabilize the phosphorylated state.  Adding KaiB to a
  dephosphorylation assay would then decrease the dephosphorylation
  speed, contrary to what is found in experiments \cite{Rust2007},
  which show no change in dephosphorylation dynamics.  In our model,
  the phosphorylation cycle only sets the timing of the conformational
  switch by regulating the activation energy for ADP dissociation in
  the CI domain.  Hydrolysis of ATP in the CI domain will continually
  generate ADP in the binding pocket of CI, but only when enough
  serine sites on CII have been phosphorylated, does the ADP release
  rate drop sufficiently so that the ADP level in CI will rise.  It is
  this rise in ADP level that stabilizes the inactive state of KaiC.
  Hydrolysis in the CI domain thus drives the conformational switch
  and provides the large change in affinity between the active and
  inactive states, necessary for KaiB binding and KaiA
    sequestration. The latter, in turn, underlies the synchronization
    of the phosphorylation cycles of the individual hexamers, which is
    essential for generating the macroscopic oscillations in
    phosphorylation level.

{\bf Positive feedback is not essential; time delay and negative feedback are sufficient.} 
Unlike in the model by Van Zon {\it et al.}, 
here the phosphorylation dynamics of the hexamer is independent for each
monomer, and a hexamer does not need to be fully phosphorylated before
flipping to the inactive state.  Phosphorylation of the threonine
and serine sites in each KaiC monomer has antagonistic effects on the
ADP dissociation rate from the CI domain, and consequently on the switch
of the conformational state.  Due to this antagonism, the
conformational switch depends on the difference of T to S phosphorylated
residues, and not on the absolute phosphorylation level.  Therefore, in
our model, a hexamer does not need to go through a full phosphorylation
cycle each period, as was the case for the model by Van Zon and coworkers.
Furthermore, there is no direct cooperativity between monomers; their
states all add linearly to the activation energy for ADP dissociation
in CI and the free energies of the conformational states.  
In particular, the D and S state have a similar effect on the ADP dissociation rate in CI, 
and hence on the conformation of the hexamer and on its ability to sequester KaiA.
This means that the synchronization mechanism of the original monomer model by
Rust et. al. \cite{Rust2007} does not apply. In their model, KaiA
prevents the occupation of the S state by enhancing the transition
from the S to the D state, while only the S state sequesters KaiA. 
This mutual inhibition between KaiA and the S state creates a positive
feedback loop for KaiA sequestration that is essential to the
oscillations in that model. 
In contrast, in our model the D and S states both stimulate KaiA sequestration, 
so it does not exhibit this positive feedback mechanism.  
Because our model acts at the level of hexamers rather than monomers, 
it does not need the positive feedback: 
the delay between the conformational switch and
the subsequent binding and sequestration of KaiA is sufficiently long
that, together with the negative feedback of KaiA sequestration on
phosphorylation, it can generate oscillations.

{\bf The model is robust to variations in ATP fraction.} 
Our model correctly reproduces the phosphorylation and dephosphorylation time
traces of the T,D and S state monomers. 
Furthermore, as in experiments \cite{Rust2011}, 
dephosphorylation is independent of the bulk ATP fraction, 
whereas phosphorylation does strongly depend on it. 
The idea that dephosphorylation is dominated by the
phosphotransfer pathway \cite{Nishiwaki2012, Egli2012} is confirmed by good agreement with the
experimentally observed ATP production in the CII domain and the delay
in the appearance of inorganic phosphate in the bulk \cite{Nishiwaki2012}.  
To compare the ATP consumption of KaiC in our model with experiments, 
we investigated the ATP fraction in the nucleotide binding pockets and the ATPase activity of KaiC,
for different mixtures of Kai proteins.
Both the transient dynamics of the ATP fraction in the binding pockets and the steady state 
ATPase rate in mixtures with KaiC only or KaiC and KaiA are in qualitative agreement with experiments.
In a system with macroscopic oscillations,
both the phase difference between the ATP fraction in the binding pockets 
and the phosphorylation fraction and the amplitude of the ATP fraction 
are in excellent agreement with experiments.  
To check the robustness of the oscillator against variations in the bulk ATP fraction, 
we checked whether oscillations persist at lower ATP fractions and if the
phosphorylation period is independent of the fraction. 
We found that the clock period is indeed constant over a wide range of ATP fractions. 

\subsubsection*{Open questions}
Although our model is able to reproduce most of the available experimental data, 
there are a few observations that it cannot replicate in its current form. 

First, in our model, KaiB barely interacts with unphosphorylated KaiC, 
in contrast with the observation that KaiB lowers the ATPase activity of a
solution with only KaiC \cite{Terauchi2007}, and that KaiB can bind
unphosphorylated WT KaiC \cite{Villarreal2013} at micromolar concentrations. 
In our model, 
it is essential that unphosphorylated KaiC is predominantly in the active conformation:
Monomers in the U state increase the dissociation rate from ADP in the CI domain, 
thereby stabilizing the active state which leads to the dissociation of the sequestered KaiA and KaiB 
at the end of the cycle. 
The consequence is that unphosphorylated KaiC is predominantly in the active conformation, 
which has a very low affinity for KaiB. 
Making the inactive state of unphosphorylated KaiC more stable 
would remedy this shortcoming of the model, because in the inactive state KaiC can bind KaiB.
Consistent with this idea, 
very recent experiments suggest that the inactive conformational state is indeed more stable: 
About half of the unphosphorylated KaiC hexamers in a system without KaiA or KaiB,
are in the inactive conformational state \cite{Oyama2016}. 
% \hl{However, in our model, oscillations stop when we increase the affinity of U-state KaiC for KaiB
% because not all sequestered KaiA is released at the end of the oscillation.
% It is therefore hard to reconcile a higher affinity for KaiB with robust oscillations 
% in our current understanding of the Kai system.}
However, increasing the affinity of KaiB for unphosphorylated KaiC also increases,
in the current model, the capacity of unphosphorylated KaiC to sequester KaiA,
which impedes the release of KaiA at the end of the cycle.
For future research, it will be interesting to see whether by amending the model 
these experimental observations can be reproduced.

Secondly, the oscillations in our model come to a standstill when the bulk ATP fraction drops below 40\%, 
while in experiments they continue to exist until the ATP fraction drops below 25\% \cite{Rust2011, Phong2012}. 
In our model, when the ATP fraction drops below a critical value of around 40\%, not
enough hexamers have 6 KaiB monomers bound to sequester all KaiA
from solution at the required phase of the oscillation. 
This impedes the synchronization of the individual hexamers, 
necessary for coherent macroscopic oscillations. 
One way to resolve this might be to increase the affinity of unphosphorylated KaiC for KaiB,
but as mentioned above, this impairs the release of KaiA at the end of the cycle.
Alternatively, or in addition, making the window of KaiA sequestration more
deterministic e.g. by making dephosphorylation of the respective
monomers within a hexamer more concerted or by more tightly coupling KaiB-KaiC binding to the KaiC
phosphorylation state, is expected to extend the range of ATP
concentration over which the model exhibits oscillations.
We have not investigated the effects of the concerted phosphorylation of hexamers,
because it deviates too much from our current model where monomers phosphorylate independently.
It is also possible that including monomer exchange might improve synchronization 
of the KaiC hexamers and thus allow oscillations to persist to lower ATP fractions, 
but such an effect likewise cannot readily be included in the current model.

Lastly, the ATP consumption in our model is slightly higher than observed.
We hypothesized that this can be attributed to the ATP
hydrolysis in the CII domain.  To provide support for this idea, 
we looked at an alternative model where
the binding of KaiA suppresses ATP hydrolysis in CII. 
This model did show ATPase rates very similar to experimentally observed values.

\subsubsection*{Predictions and experimental verification}
Here we explore the possibilities for experimentally verifying the predictions from our model.
Our model is based on two important ingredients: 1) The relative stability of the two
conformations is determined by the ATP fraction in the binding pockets of the CI domain and
2) this fraction is set by the number of phosphorylated serine 
and threonine sites in the CII domain of the hexamer.

The dependence of the conformation on the nucleotide binding state of CI can be 
tested by measuring the ATP fraction in the binding pockets of KaiC \cite{Terauchi2007, Nishiwaki-Ohkawa2014}
while at the same time probing the conformational state as in the study of \cite{Oyama2016},
where the authors track the fractions of active and inactive KaiC hexamers over time.
Our analysis predicts a positive correlation between the fraction of ADP bound
to CI and the fraction of inactive KaiC,
including KaiC mutants with a different hydrolysis rate constant in CI \cite{Terauchi2007} 
and in an assay where KaiC is in the presence of KaiA and KaiB, and oscillates over time.
% The positive correlation between the ATPase activity and the oscillation frequency 
% as observed in \cite{Terauchi2007,Abe2015}, can in part be explained by our model.
% \hl{A higher hydrolysis rate in the CI domain will increase the ADP level in CI,
% }

The dependence of the ATP fraction in the binding pockets of the CI domain 
on the phosphorylation state of the CII domain
can be tested by measuring the concentrations of monomers in the 
U,T,D and S phosphorylated state \cite{Rust2007, Murayama2011}
and again the ATP fraction in the nucleotide binding pockets \cite{Terauchi2007, Nishiwaki-Ohkawa2014}.
We predict that serine-phosphorylated KaiC slows down the dissociation of ADP from CI,
decreasing its ATP fraction in the binding pocket,
and threonine-phosphorylated KaiC should antagonize this effect.
Starting with different ratios and levels of monomers phosphorylated at their threonine and serine sites,
either using KaiC phosphomimics or aliquots from an oscillating system,
the ADP fraction in the CI binding pockets should show 
a positive correlation with the \emph{difference} between 
the number of phosphorylated serine sites and threonine sites.
Adding KaiB should enhance the effect, 
because it cooperatively stabilizes the inactive state with ADP via the MWC mechanism. 
Our model can also explain the observation described in \cite{Abe2015}, 
where they found that adding ATP to unphosphorylated KaiC 
leads to a transient dip in the ATPase rate:
The transient phosphorylation of KaiC temporarily lowers the CI-ADP dissociation rate. 
Clearly, it would be of interest to repeat these
experiments starting with KaiC in different phosphorylation states, 
and in the presence and absence of KaiB. 
Our model predicts that, starting from fully phosphorylated KaiC, 
during dephosphorylation the ATP fraction in the binding pockets will exhibit a dip (see \fref{CICII_ATPase}).

Related to this, and more specifically, our model predicts that the ADP dissociation rate 
in the CI domain is set by the relative number of phosphorylated serine and threonine sites
in the CII domain, and not by their absolute levels.
This implies that not all the residues have to be phosphorylated before a hexamer
can switch to the inactive conformation and complete its cycle.
Indeed as we show in a forthcoming publication, at lower ATP fractions of the buffer,
hexamers go through a smaller phosphorylation cycle.
This could in principle be tested experimentally if it were possible to track individual hexamers 
as they go through their phosphorylation cycle.

While these experiments test our predictions on the connection between the CI and CII domain,
our analysis also predicts an interesting consequence of the idea that phosphotransfer is the major 
pathway for dephosphorylation \cite{Nishiwaki2012, Egli2012}. 
This could be tested by revisiting the experiments on
dephosphorylation of radioactively labeled KaiC \cite{Nishiwaki2012}, 
but now in solution with non-hydrolyzable ATP.
Since the ATP can not be hydrolyzed, there will be no ADP in the CII
binding pockets, and the phosphate groups on the S and T sites in KaiC
can not be transferred to ADP in the CII domain.  This should
significantly slow down the dephosphorylation speed if indeed
phosphotransfer is the dominant pathway.

Our model of the interaction of KaiA with the CI and CII domains of KaiC 
also allows us to make predictions for how long KaiA is bound to one of the domains 
during an oscillation.
In our model, the dissociation rate of KaiA from the CII domain is much higher than
the frequency of the oscillation.
This is essential for differential affinity,
where KaiA continually binds different KaiC hexamers during the phosphorylation phase 
to promote the phosphorylation of hexamers that are lagging behind.
Our model therefore predicts a sharp peak in the distribution
of times for which the CII domain of a certain KaiC hexamer is bound to KaiA during an oscillation period,
$\Delta\rmlabels{t}{}{CII\cdot KaiA}$.
Indeed, \fref{HistKaiASeq}A shows a clear single peak in this distribution.
When we lower the dissociation rate of KaiA from the CII domain,
this distribution broadens, \fref{HistKaiASeq}B, indicating differential affinity is hampered.
One might think that when the dissociation rate is decreased even further,
the distribution in $\Delta\rmlabels{t}{}{CII\cdot KaiA}$ becomes bimodal,
because once KaiA is bound to the CII domain of a certain hexamer, 
it will continue to stay bound to this hexamer during the whole phosphorylation phase
of an oscillation cycle; 
because KaiA is limiting, this means that other KaiC hexamers will not, via their CII domain, 
bind KaiA during that cycle.
In this case one fraction of hexamers does not or only very briefly binds KaiA via the CII domain
while the other fraction is bound to KaiA for most of the time during the phosphorylation phase.
However, in our model oscillations stop when we decrease the dissociation rate of KaiA 
to such a low level to allow for a bimodal distribution.
Nevertheless, bi-modality could arise in experiments, 
which would indicate that the role of KaiA during the phosphorylation phase 
is very different from what we predict in our model.
For KaiA bound to the CI domain, our simulations show exponentially 
distributed bound times, \fref{HistKaiASeq}B.
This distribution is bimodal, as only 70\% of the hexamers sequester KaiA,
while the other 30\% do not make it to the inactive state and bind six KaiB during an oscillation.
Techniques to follow protein complex formation at the single molecule level have been developed \cite{Viani2000}, 
suggesting that future experiments might be able to reveal how long KaiA is bound to KaiC
during an oscillation cycle.
% Future experiments should be able to reveal how long KaiA is bound 
% to the domains of KaiC during an oscillation cycle.

%100 ATP, std,       CI:0.70 occupied, CII: 0.99
% 50 ATP, std, 	     CI:0.60 occupied, CII: 1.00
%100 ATP, NoDiffAff, CI:0.75 occupied, CII: 0.99
\begin{figure}[ht!]
\begin{center}
\includegraphics[scale=1.0]{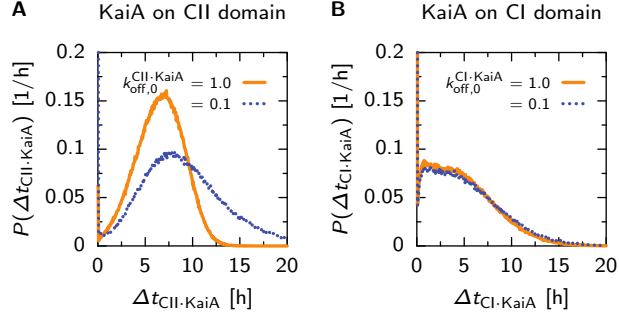} 
\end{center}
\caption{\flabel{HistKaiASeq} Probability density of the time, 
per period of the phosphorylation cycle,
the CII domain of a KaiC hexamer is bound to KaiA, $\Delta\rmlabels{t}{}{CII\cdot KaiA}$ (A),  
or the CI domain has at least one KaiA dimer bound, $\Delta\rmlabels{t}{}{CI\cdot KaiA}$ (B).
We compare situations with our standard value for the KaiA dissociation rate,
$\rmlabels{k}{CII\cdot KaiA}{off, 0}=1.0$ (orange solid line) with
a much lower value $\rmlabels{k}{CII\cdot KaiA}{off, 0}=0.1$ (blue dotted line).
(A) For the CII domain, both values of the KaiA dissociation rate show unimodal distributions 
of the time KaiA is bound.
This indicates that, during a period, KaiA binds to all hexamers in the ensemble equally likely.
Of all hexamers, only 1\% do not bind KaiA to CII at any time during the period.
(B) For the CI domain, the time KaiA is sequestered by KaiC is roughly exponentially distributed,
and is the same for both KaiA dissociation rates from the CII domain.
Here, 30\% of the hexamers do not sequester KaiA at all during a period,
indicating many hexamers do not sequester KaiA during a cycle.}
\end{figure}

\section*{Methods}
We model the post\hyp{}translational Kai oscillator using kinetic Monte Carlo \cite{Gillespie1977, Gillespie2007}.
Each monomer has two nucleotide binding states for both the CI and CII domain and 4
phosphorylation states. Additionally, a hexamer can be in an active or inactive state.
This results in $(2\cdot 2\cdot 4)^6\cdot 2=2^{25}$ different states a single hexamer can be in.
Furthermore, there are the binding reactions with the KaiC hexamer: 
2 for KaiA with the CII domain, 
6 for KaiB monomers to the CI domain and another 6 reactions for KaiA binding to the CI domain of 
KaiC with 6 KaiB monomers.
The total number of reactions therefore exceeds a billion. 
Clearly, the reaction combinatorics makes a straightforward ODE or Gillespie simulation unfeasible.
We thus set out to design a dedicated kinetic Monte Carlo algorithm to simulate the Kai system,  
where we do not have to write down all possible reactions explicitly. 

To this end, we designed an algorithm that keeps track of
\rmlabels{N}{hex}{} KaiC hexamers, which in turn consist of six
explicitly simulated monomers, and \rmlabels{N}{KaiA}{tot} KaiA dimers.
Since we have defined individual hexamers and monomers, we can
calculate the propensity that a reaction occurs in a particular
hexamer, and, in turn, the propensity that a reaction occurs in a
monomer that is part of this hexamer.  This allows us to create a
layered version of the original algorithm by Gillespie
\cite{Gillespie1977}, where we first determine the next reaction time
and in which hexamer this reaction will take place.  Then, in the next
step, we determine which reaction or monomer of this hexamer fires.
If a monomer fires, we choose which reaction of this monomer happens.
This layered approach allows us to separate state changes in a hexamer
that only modify the reaction propensities of that specific hexamer from
state changes that influence the KaiA concentration in solution, which
affects all hexamers.  This greatly reduces the reaction combinatorics
and the computational cost of the algorithm.

Specifically, the state of the whole system, \brmlabels{s}{tot}{}, consists of the state of the 
hexamers with index $h$, \brmlabels{s}{hex}{h}, and the number of KaiA dimers in solution, \newline\rmlabels{n}{KaiA}{sol}:
$\brmlabels{s}{tot}{}=\{\{\brmlabels{s}{hex}{h}\}_{h=1}^{\rmlabels{N}{hex}{}}, \rmlabels{n}{KaiA}{sol}\}$.
The hexamer state contains the state vectors of its six monomers with index $m$, \brmlabels{s}{mon}{h,m}, 
the conformational state, $C$, the number of KaiA bound to CI, \rmlabels{n}{CI\cdot KaiA}{}, 
the number of KaiB bound to CI, \rmlabels{n}{CI\cdot KaiB}{}, 
and the number of KaiA bound to the CII domain, \rmlabels{n}{CII\cdot KaiA}{}:
$\brmlabels{s}{hex}{h}=\{\{\brmlabels{s}{mon}{h,m}\}_{m=1}^{6}, C, \rmlabels{n}{CI\cdot KaiA}{},\rmlabels{n}{CI\cdot KaiB}{}, \rmlabels{n}{CII\cdot KaiA}{}\}_{h}$.
The state vector of a monomer consists of the threonine and serine phosphorylation site 
and the nucleotide binding pockets of the CI and CII domain:
$\brmlabels{s}{mon}{h,m}=\{S,T,\rmlabels{n}{CI}{nucl},\rmlabels{n}{CII}{nucl}\}_{h,m}$.

Given the system state, we can calculate the firing propensity of
reaction $\mu$ that chang\-es the state vector of monomer $m$,
which is part of hexamer $h$, \rmlabels{q}{\mu}{h,m}(\brmlabels{s}{hex}{h}).
Note that this reaction propensity can depend on the state of the whole hexamer,
and therefore is a function of the hexamer state vector, and not only of the state vector of monomer $m$.
The propensity for firing a reaction with index $\nu$, which changes the state variables of hexamer $h$,
denoted \rmlabels{q}{\nu}{h}(\brmlabels{s}{hex}{h}, \rmlabels{n}{KaiA}{sol}),
depends on the state vector of hexamer $h$ only, and the number of KaiA in solution.
Given these reaction propensities, we can calculate the accumulated propensities, denoted by $\tilde{q}$,
of firing a single monomer $m$ in hexamer $h$, a single hexamer $h$ and the total propensity as
\begin{eqnarray}
 \rmlabels{\tilde{q}}{mon}{h,m}(\brmlabels{s}{hex}{h}) & = & \sum_{\bf \mu} \rmlabels{q}{\mu}{h,m}, \\
 \rmlabels{\tilde{q}}{hex}{h}(\brmlabels{s}{hex}{h}, \rmlabels{n}{KaiA}{sol}) & = & \sum_{m=1}^{6} \rmlabels{\tilde{q}}{mon}{h,m} + \sum_{\bf \nu} \rmlabels{q}{\nu}{h}, \\
 \rmlabels{\tilde{q}}{tot}{}(\{\brmlabels{s}{hex}{}\}^{\rmlabels{N}{hex}{}}_{h=1}, \rmlabels{n}{KaiA}{sol}) & = & \sum_{h=1}^{\rmlabels{N}{hex}{}} \rmlabels{\tilde{q}}{hex}{h},
\end{eqnarray}
respectively.

Since the firing of a reaction is a Markov process, 
the probability of hexamer $h$ firing in the infinitesimal time interval $[t+\tau,t+\tau+d\tau]$ is
\begin{equation}
 P(\tau,h|\brmlabels{s}{tot}{}, t)d\tau= \rmlabels{\tilde{q}}{hex}{h}\,{\rm exp}\left( -\rmlabels{\tilde{q}}{tot}{}\tau \right)d\tau.
\end{equation}
Now, given two random numbers, $\rho_1$, $\rho_2$, drawn from a uniform distribution with domain $[0,1)$, 
we calculate the next event time, $\tau$, and the hexamer to fire, $h$, as
\begin{eqnarray}
 \tau & = & \frac{1}{\rmlabels{\tilde{q}}{tot}{}}{\rm ln}\left( \frac{1}{\rho_1} \right), \elabel{Gill_tau} \\
 h & = & \text{the smallest integer satisfying}\,\sum_{h'=1}^{h} \rmlabels{\tilde{q}}{hex}{h} > \rho_2 \rmlabels{\tilde{q}}{tot}{},
 \elabel{Gill_index}
\end{eqnarray}
respectively.

Having defined the important propensities, our dedicated kinetic Monte Carlo algorithm becomes
\begin{enumerate}
 \item[0.] Initialize the time $t=t_0$ and the system's state $\brmlabels{s}{tot}{}=\brmlabels{s}{tot}{0}$. 
           Calculate the propensities \rmlabels{q}{\mu}{h,m}, \rmlabels{q}{\nu}{h}, \rmlabels{\tilde{q}}{mon}{h,m}, 
           \rmlabels{\tilde{q}}{hex}{h} and \rmlabels{\tilde{q}}{tot}{}.
 \item[1.] Calculate the time interval to the next reaction, $\tau$, using \eref{Gill_tau}.
 \begin{enumerate}
  \item[a.] Choose which hexamer, $h$, to fire, with $P(h|\tau)=\rmlabels{\tilde{q}}{hex}{h}/\rmlabels{\tilde{q}}{tot}{}$.
  \item[b.] Choose which reaction, $\nu$, with $P(\nu|\tau,h)=\rmlabels{q}{\nu}{h}/\rmlabels{\tilde{q}}{hex}{h}$ 
            or monomer, $m$, with $P(m|\tau,h)=\rmlabels{q}{mon}{h,m}/\rmlabels{\tilde{q}}{hex}{h}$, to fire.
  \item[c.] If a monomer was chosen,
	    choose which reaction to fire with $P(\mu|\tau,h,m)=\rmlabels{q}{\mu}{h,m}/\rmlabels{\tilde{q}}{mon}{h,m}$.
 \end{enumerate}
  \item[2.] Fire reaction and update the state vector of hexamer $h$, and \rmlabels{n}{KaiA}{sol} 
            in case of a bimolecular reaction.
  Recalculate all reaction propensities \rmlabels{q}{\nu}{h} and \rmlabels{q}{\mu}{h,m} for each monomer $m$.
  In case a bimolecular reaction was fired, change $\rmlabels{n}{KaiA}{sol}$ accordingly, and update 
  the bimolecular reaction propensities in all hexamers.
  \item[3.] Recalculate \rmlabels{\tilde{q}}{mon}{h,m} and \rmlabels{\tilde{q}}{hex}{h} for the fired hexamer $h$.
  In case of a bimolecular reaction, also recalculate \rmlabels{\tilde{q}}{hex}{h'} for all hexamers $h'$.
  Update \rmlabels{\tilde{q}}{tot}{}. 
  \item[4.] Record ($t, \brmlabels{s}{tot}{}(t)$) as desired, return to step 1.
\end{enumerate}

\section*{Acknowledgments}
  %We thank TODO for a critical reading of the manuscript. 
  This work was supported in part by FOM, 
  which is financially supported by the Nederlandse Organisatie voor Wetenschappelijk Onderzoek
  (JP and PRtW), and by NSF Grant DMR-1056456 (DKL).

% TEMPORARILY FOR GENERATION OF BBL FILE
% \bibliographystyle{plos2015}
% \bibliography{library}

\end{document}